\documentstyle[psfig]{mn}

\newcommand{\kal}[1]{\begin{cal} #1 \end{cal}}
\newcommand{\bgq}[0]{$B_{\rm gq}\ $}
\newcommand{\agq}[0]{$A_{\rm gq}\ $}
\newcommand{\n}[0]{$N_{0.5}\ $}
\newcommand{\lsim}{\raise0.3ex\hbox{$<$}\kern-0.75em{\lower0.65ex\hbox{$\sim$}}}
\newcommand{\gsim}{\raise0.3ex\hbox{$>$}\kern-0.75em{\lower0.65ex\hbox{$\sim$}}}

\title{Clustering of galaxies around radio quasars at $0.5\leq z \leq0.8$}

\author[Wold et al.\ ]
 {Margrethe Wold,$^{1,3}$
  Mark Lacy,$^2$
  Per B.~Lilje$^3$ and 
  Stephen Serjeant$^4$\\
  $^1$Stockholm Observatory, SE-133 36 Saltsj{\"o}baden, Sweden \\
  $^2$Astrophysics, Nuclear and Astrophysics Laboratory, Keble Road, Oxford, OX1 3RH, U.K. \\
  $^3$Institute of Theoretical Astrophysics, University of Oslo, P.O. Box 1029 Blindern, N-0315 Oslo, Norway \\
  $^4$Astrophysics Group, Imperial College London, Blackett Laboratory, Prince Consort Road, London SW7 2BZ, U.K. \\}

\date{Accepted 0000.
      Received 0000;
      in original form 0000}

\pubyear{1999}

\begin{document}

\maketitle

\begin{abstract}
We have observed the galaxy environments around a sample of 21 radio-loud, steep-spectrum
quasars 
at $0.5\leq z \leq0.82$, spanning several orders of magnitude in radio luminosity.
The observations also include background control fields used to obtain the excess number
of galaxies in each quasar field.
The galaxy excess was quantified using the spatial galaxy--quasar correlation 
amplitude, $B_{\rm gq}$, and an Abell-type measurement, $N_{0.5}$ (Hill \& Lilly 1991). 
A few quasars are found in relatively rich clusters, but on average, they 
seem to prefer galaxy groups or clusters of approximately Abell class 0. 
We have combined our sample with literature samples extending down to $z\approx0.2$ 
and covering the same range in radio luminosity. By using 
Spearman statistic to disentangle redshift and luminosity dependences, we 
detect a weak, but significant, positive correlation between the richness of the 
quasar environment and the quasar's radio luminosity. 
However, we do not find any epoch dependence in $B_{\rm gq}$, as has previously been reported 
for radio quasars and galaxies.   
We discuss the radio luminosity--cluster richness link and possible explanations for
the weak correlation that is seen. 
\end{abstract}


\section{Introduction}
\label{section:section1}

Since the discovery of quasars, much effort has been expended in trying
to understand the physics of active galactic nuclei (AGN). Most work
has concentrated on their nuclei, but during the last 10--15 years the galaxy 
environments around different classes of AGN have received much attention.

The first studies that were made in this field (e.g.\ Stockton 1978; Longair \& Seldner
1979; Yee \& Green 1984; Yee \& Green 1987 (hereafter YG87); Prestage \& Peacock 1988, 1989; 
Yates, Miller \& Peacock 1989) showed that radio quasars
and galaxies are located in regions of enhanced galaxy density.
These findings were confirmed by more recent studies (e.g.\ Ellingson, Yee \& Green 1991
(hereafter EYG91); Hill \& Lilly 1991; Allington--Smith et al. 1993; Fisher, Bahcall \& 
Kirhakos 1996; Zirbel 1997), which
added more weight to the consensus that the very local environment may be important for 
the formation and evolution of AGN.
It is believed that AGN formation is closely related to the formation of 
supermassive black holes, possibly by galaxy--galaxy interactions and
mergers. Interactions and mergers may also provide the fuel supply to a
pre-existing central black hole. The study of AGN environments may therefore
provide important clues about their formation and evolution.

Radio-loud AGN are found to be associated almost exclusively
with giant elliptical host galaxies (e.g.\ Taylor et al.\ 1996; Lacy et al.\ 1999
and references therein). Both the need for
interaction, and the apparent need for a massive host, make it not surprising
that groups or clusters of galaxies might be associated with radio-loud
AGN. Furthermore, because the efficiency of conversion of the bulk
kinetic energy of the radio jets to radio luminosity increases with 
environmental gas density, selection effects will also tend to favour
finding steep-spectrum radio sources in gas-rich environments, i.e.\ 
in groups or clusters. High-$z$ radio-loud AGN can therefore be used 
as tracers of groups and clusters at early epochs, and the evolution 
of galaxy groups and clusters can thus be studied independent of the selection
biases associated with optical or X-ray cluster surveys.

Studying AGN environments may also help establish the 
relationship between different classes of AGN. In particular it has been
proposed that radio galaxies and radio-loud quasars are the same type
of object, but viewed at different orientations to the line of sight 
(Scheuer 1987; Barthel 1989; Antonucci 1993). The most robust test of 
such orientation-based `unified models', is to select the AGN 
in question with one orientation-independent quantity (e.g.\ radio lobe
luminosity) and compare another (e.g.\ environment).
And indeed, observations of fields around radio-loud 
quasars and galaxies show that they are sited in environments that occupy approximately 
the same range in galaxy density (e.g.\ Longair \& Seldner 1979; Smith \& Heckman 1990;
Hill \& Lilly 1991; Yee \& Ellingson 1993).

For obvious reasons, most studies have been concerned with sources at
$z\lsim0.6$, but recently various lines of evidence point to radio-loud AGN being in
cluster environments also at $z\gsim1$ (e.g. Hall \& Green 1998; Dickinson 1996). 
Complications due to the possibility of lensing bias (Hammer \& Le~F\`{e}vre 1990; 
Fort et al.\ 1996; Ben\'{\i}tez, Mart\'{\i}nez-Gonz\'{a}lez \& Mart\'{\i}n-Mirones 1997;
Schneider et al.\ 1998), the possibility of non-thermal
contributions to X-ray emission, and the low contrast of clusters against
the background, however, make detections of clusters at these redshifts hard to prove.

An apparent decrease in the galaxy density around powerful (FRII) quasars from 
$z\sim0.5$ to the present epoch has been observed (YG87, EYG91), although the amount 
of evolution is still not firmly established. 
Several investigators have also found that powerful FRII radio galaxies
inhabit richer galaxy clusters at $z\sim0.5$ than at $z\lsim0.3$. 
On the other hand, the less powerful FRI galaxies seem to be commonly found 
in clusters at both higher and lower redshifts (Longair \& Seldner 1979; Prestage \& Peacock 
1988, 1989; Hill \& Lilly 1991; Zirbel 1997).
Yates et al.\ \shortcite{yates} studied the environments of powerful
radio galaxies in a wide redshift range and found that sources at $z>0.3$ 
reside in clusters of typically Abell \cite{abell} richness class 0, while the 
environments at $z<0.3$ are three times poorer.
Hill \& Lilly \shortcite{hill} compared a sample of radio galaxies at
$z\approx0.5$ spanning a wide range in radio luminosity with a lower redshift 
sample by Prestage \& Peacock (1988, 1989) and confirmed the Yates et al.\ result.
Allington--Smith et al.\ \shortcite{allington-smith} also arrived at a similar result, 
but with a less strong correlation between redshift and richness.

Most of the samples that have been studied are flux-limited, i.e.\ they suffer from 
a strong correlation between redshift and radio luminosity. This raises the 
possibility that the observed evolution in richness is dependent on radio luminosity
rather than redshift. Yates et al.\ (1989) believed that 
this was the case in their study, but they were unable to confirm it since their
sample had this strong redshift-radio luminosity correlation. 
However, for radio galaxies, 
Hill \& Lilly \shortcite{hill} were able to show that this is not the case by 
observing $z\sim 0.5$ sources over a wide range in 
radio luminosity. They found no correlation between richness
and radio luminosity within their sample.

In this paper, we investigate the situation at 
$0.5 \leq z \leq 0.82$ for radio-loud quasars.
This redshift range was chosen to extend 
previous work, which went up to $z\sim 0.6$, to as high a redshift 
as possible consistent with keeping the redshifted 4000 {\AA} break 
shortward of the $I$-band. We selected quasars randomly from flux-limited 
samples spanning as wide a range in radio luminosity as possible. 
By extending  
the redshift range of previous studies, whilst maintaining the luminosity 
range, we are able to better disentangle trends in environmental
richness due to cosmic evolution from those due to radio luminosity.
The work presented here is also part of a larger study which aims at
investigating and comparing the galaxy environments of different 
classes of AGN within a relatively small redshift range, 
but over as wide a luminosity range as possible.

We quantify the amount of galaxy clustering in each quasar field
using the spatial quasar--galaxy cross-correlation amplitude, 
$B_{\rm gq}$, and an Abell-type measurement, $N_{0.5}$. 
These two quantities are briefly described in Section~\ref{section:section2}.
In Section~\ref{section:section3} and \ref{section:section4},  
we present the quasar sample and describe the observations.
The data reduction, photometry and faint-object
detection in the CCD images are discussed
Section~\ref{section:section5}.
In Section~\ref{section:section6}, we quantify the galaxy environments in 
the quasar fields using \bgq and $N_{0.5}$, and compare our results with
previous work on radio-loud AGN and other two--point
correlation functions. In Section~\ref{section:section7}, we investigate 
redshift and luminosity dependences in \bgq in our own sample, and in a 
combined sample 
where our sources are put together with radio-loud quasars from other studies.
We discuss our findings in 
Section~\ref{section:section8}, and the results are
summarized in Section~\ref{section:section9}. 

Unless otherwise specified, we have assumed an Einstein--deSitter universe model 
with $H_{0}=50$ km s$^{-1}$ Mpc$^{-1}$ and $q_{0}=0.5$ throughout the analysis. 
Also, in this paper, `quasar' should be interpreted as
`radio-loud quasar'.

%************************************************************************

\section{Quantifying galaxy excess}
\label{section:section2}

\subsection{Galaxy--quasar cross-correlation amplitude}

The use of the amplitude of the spatial cross-correlation function as a
measure of galaxy excess was initiated by Seldner \& Peebles 
\shortcite{seldner} and Longair \& Seldner \shortcite{longair}.
Since then, it has been
widely used by other investigators for quantifying galaxy environments around
various types of AGN at both lower and higher redshifts
(e.g. Yee \& Green 1984; YG87; Prestage \& Peacock 1988, 1989; Yates et al.\ 1989; EYG91; 
Smith, O'Dea \& Baum 1995; Wurtz et al.\ 1997; De~Robertis, Yee \& Hayhoe 1998).
We only give the expressions here, but a detailed derivation 
may be found in Longair \& Seldner \shortcite{longair}.

The spatial cross-correlation function, $\xi\left(r\right)$, is found to 
be well fit by a power--law with slope $\gamma$ and amplitude $B_{\rm gq}$,
\[\xi\left(r\right)=B_{\rm gq}r^{-\gamma},\]
\noindent
where $B_{\rm gq}$ is the amplitude we seek in this analysis.
In order to determine \bgq from the data, the amplitude, $A_{\rm gq}$, of the {\em angular}
cross--correlation function, $\omega(\theta)=A_{\rm gq}\theta^{1-\gamma}$, must first
be determined. An estimator for this amplitude is:
\begin{equation}
A_{\rm gq}=\frac{N_{\rm tot}-N_{\rm b}}{N_{\rm b}}\frac{3-\gamma}{2}\,\theta^{\gamma -1},
\label{eq:equation1}
\end{equation}
\noindent
which is directly obtainable from galaxy counts in the quasar fields
and in images of background control fields.
In the above expression, $N_{\rm tot}$ is the number of galaxies in the quasar 
field within a circle of radius $\theta$ centered on the quasar,
corresponding to 0.5 Mpc at the quasar redshift. The number of 
background galaxies from the control fields 
is denoted by
$N_{\rm b}$. The slope of the correlation function, $\gamma$, is 
usually assumed to be 1.77, as found by Groth \& Peebles \shortcite{groth} for the 
low-redshift galaxy--galaxy correlation function. 
YG87 showed that the galaxy--quasar
correlation function has a slope consistent with this. 
Lilje \& Efstathiou \shortcite{lilje}, however, found $\gamma=2.2$ in a study
of the cross-correlation between Abell clusters and the Lick galaxy counts
down to scales of 0.2 Mpc, but Prestage \& Peacock (1988, 1989) have
demonstrated that \bgq is insensitive to an incorrect choice of
$\gamma$, provided that $\gamma\sim2$. We have therefore made the usual
assumption that $\gamma=1.77$.

The spatial correlation amplitude is obtained from \agq by the relation:
\[B_{\rm gq}=\frac{\kal{N}_{\rm g}A_{\rm gq}}{\Phi\left(m_{\rm lim},z\right)I_{\gamma}}\,d_{\theta}\,^{\gamma-3},\]
\noindent
where $\kal{N}_{\rm g}$ is the 
average surface density of background galaxies, and 
$d_{\theta}$ is the angular diameter distance to the quasar.
The two quantities in the denominator are $I_{\gamma}=3.78$, which is an integration 
constant, and $\Phi\left(m_{\rm lim},z\right)$, which denotes a universal luminosity 
function integrated down to the apparent completeness limit at the quasar redshift: 
$\Phi\left(m_{\rm lim},z\right)=
\int_{L\left(m_{\rm lim},z\right)}^{\infty}{\phi\left(L\right){\rm d}L}$.
The integrated luminosity function gives the number of field galaxies per unit volume
at the quasar redshift with magnitudes brighter than the completeness limit
of the data. By normalizing \bgq with $\Phi\left(m_{\rm lim},z\right)$, an absolute
measure of the galaxy excess at the quasar redshift is obtained, 
but unfortunately the luminosity function is poorly
constrained at higher redshifts. This a critical step in the analysis, especially since \bgq is 
inversely proportional to the normalization, $\phi^{*}$, of this luminosity function.
As advocated by 
YG87 and Yee \& L{\'o}pez--Cruz \shortcite{yeelopez}, it is therefore important that $\phi^{*}$
is consistent with the background galaxy counts obtained from the data.
In our analysis, we have used the Canada--France Redshift Survey (CFRS) luminosity function
\cite{lilly} to track the evolution of the characteristic luminosity $L^{*}$, and then used the  
observed galaxy counts in order to fit $\phi^{*}$ as a function of redshift.
This is described in more detail in Section~\ref{section:section6}.

\subsection{A more direct measurement}

The quantity $N_{0.5}$, first introduced by Hill \& Lilly \shortcite{hill},
provides a more direct, but less deep measure of the clustering strength
than $B_{\rm gq}$. The galaxy counting is restricted to a certain magnitude
interval, but a separate calculation of the luminosity function is not required.
Hill \& Lilly used it for quantifying galaxy environments
around a sample of radio galaxies.
For a radio galaxy with apparent magnitude $m_{\rm g}$, \n is the
number of excess galaxies with magnitudes in the range $m_{\rm g}$ to 
$m_{\rm g}+3$ within a radius of 0.5 Mpc centred on the radio galaxy.
Since the quasar appears as a bright stellar object, the magnitude of the
host galaxy cannot be directly
determined. Instead, we have
estimated $m_{\rm g}$ for the quasar host 
from the magnitude--redshift relation for radio galaxies of Eales \shortcite{eales}:
\begin{equation}
m_{\rm g}\left(R\right)=21.05+5.3\log z, 
\label{eq:equation2}
\end{equation}
\noindent
valid for $19.0<m_{\rm g}\left(R\right)<21.3$. Some of our
quasar fields were also imaged in $I$, and in this case we estimated 
$m_{\rm g}\left(I\right)$ by subtracting the $R-I$ 
colour of an E/S0 galaxy model \cite{guiderdoni} from 
$m_{\rm g}\left(R\right)$. 
The $N_{0.5}$ measurement may suffer from systematic errors with redshift
if radio galaxies and clusters galaxies evolve differently in luminosity.
At these redshifts, however, the effect is likely to be small, and we 
have therefore not taken it into account.

\begin{figure}
\psfig{file=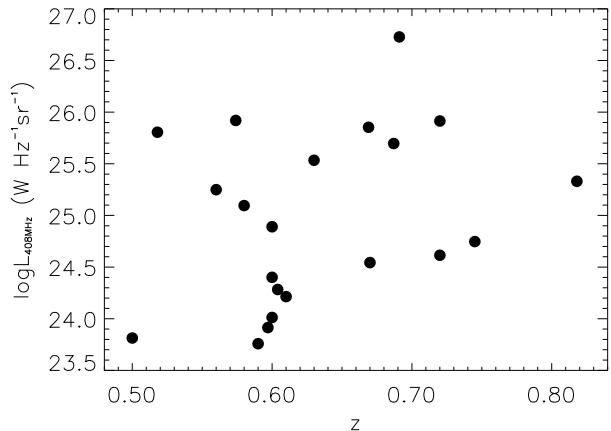}
\caption{Radio luminosity at 408 MHz of quasars in the sample as a function of redshift.
The luminosities were calculated from flux densities and spectral 
indices given in Table~1.}
\label{fig:figure1}
\end{figure}

%***********************************************************

\section{Sample selection}
\label{section:section3}

The sample investigated here consists of 21 radio-loud steep-spectrum
($\alpha>0.5$ where radio flux density is 
$S_{\nu}\propto\nu^{-\alpha}$)
quasars with redshifts $0.5 \leq z \leq 0.82$ covering the radio luminosity
range $23.8\leq \log\left(L_{\rm 408MHz}/{\rm W Hz^{-1} sr^{-1}}\right) \leq26.7$.
The sources were divided into high- and low radio luminosity
sub--samples selected from two different radio/optical flux-limited
samples in order to better distinguish between luminosity and redshift
dependences.

The eleven sources in the low-luminosity sample were
drawn from the 7C quasar (7CQ) survey (Riley et al.\ 1999) which consists of 7C sources
with a flux density limit of $S_{\rm 151MHz}>0.1$ Jy ($\approx$ 0.05 Jy at 408 MHz)
identified with objects on POSS--I (Palomar Observatory
Sky Survey) $E$-plates with $E<20$ and
colour $O-E<1.8$ ($E\simeq R$ and $O\simeq B$).
Consequently, the 7CQ survey finds only the optically
brightest steep-spectrum quasars at any redshift.
Willott et al.\ \shortcite{willott} estimate the 7CQ survey to be 40 per cent 
complete due to the bright optical magnitude limit. However, at $z=0.5$ and 
$z=0.8$, the optical flux limit corresponds to $M_{B} = -21.7$ and
$M_{B} = -22.7$ respectively (for $K$-corrections, see below), 
hence according to the definition employed by Willott et al., where a quasar must have 
$M_{B} < -23$, the 7CQ sample is complete in our redshift range. Clearly
though, this is a rather artificial distinction, and we will be missing many
broad-line objects with fainter nuclei. 

\begin{table*}
\begin{minipage}{14.5cm}
\caption{Basic data of the sources in the radio-loud quasar sample. The spectral 
index, $\alpha$, is defined in the sense that radio flux density is $S_{\nu}\propto \nu^{-\alpha}$. 
The data on the MRC and 7C quasars are from Serjeant \protect\shortcite{serjeant} and
Riley et al.\ \protect\shortcite{riley}, respectively.  
Redshift and flux density of 5C6.189 are from Rossitter \protect\shortcite{rossitter} and the redshift of 
3C380 from Laing et al.\ \protect\shortcite{laing}. The flux densities at 408 MHz of the 7C quasars were 
calculated from flux densities at 151 MHz, assuming spectral indices as given in the table.}
\begin{tabular}{lllllll}

Source& $z$ & $M_{B}$ & $S_{\rm 408}$ & $\alpha$ &  Radio       & Radio \\  
      &     &   &          (Jy)   &                   & size (kpc)   & morphology\footnote{The radio 
sizes and morphologies were determined from maps by Kapahi et al.\ (1998) 
(MRC 0032$-$203 and 0106$-$233), 
Morganti, Killeen \& Tadhunter (1993) (MRC 0159$-$117 and 0405$-$123), 
Rossitter (1987) (5C6.189), Downes et al.\ (1986) (MRC 0222$-$008), 
Reid et al.\ \shortcite{reid} (3C380) and Riley et al.\ (1999)
(7C sources). An L-band VLA map of MRC 0406-180 is presented in Best, 
R\"{o}ttgering \& Lehnert (1999) where it is consistent with being unresolved,
and it is used as a VLA phase calibrator at L-band in A-array, so must have an angular size 
$\lsim0.5$ arcsec.
Radio sizes for MRC 0033$-$000 and 0144$-$058 are from Blundell 
(private communication).}\\
      &     &    &                &               &           & \\
MRC 0032$-$203 & 0.518 & $-$21.48 & 6.870 & 1.10  & 21.7 & CSS (borderline)\\
MRC 0033$-$000 & 0.560 & $-$22.44 & 1.790 & 0.86  & 29.4 & CSS (borderline)\\
MRC 0106$-$233 & 0.818 & $-$23.96 &  1.130 & 0.60 & 20.7 & CSS (borderline)\\
MRC 0144$-$058 & 0.630 & $-$23.30 & 2.880 & 0.72  & 52.6 & FRII\\
MRC 0159$-$117 & 0.669 & $-$24.88 & 5.700 & 0.58  & 118.4 & FRII\\
5C6.189        & 0.597 & $-$23.93 & 0.091 & 0.36  & $<$ 22.9 & compact\\
MRC 0222$-$008 & 0.687 & $-$23.28 & 3.150 & 0.92  & 106.5 & FRII\\
MRC 0405$-$123 & 0.574 & $-$24.86 & 8.170 & 0.80  & 260 & FRII\\
MRC 0406$-$180 & 0.720 & $-$22.29 & 5.600 & 0.60  & $<$ 4 & CSS \\
7C2671         & 0.745 & $-$22.97 & 0.340 & 0.68  & 341.0 & FRII\\
7C2676         & 0.604 & $-$25.80 & 0.202 & 0.43  & 191.2 & FRII\\
7C2704         & 0.580 & $-$23.04 & 1.232 & 0.74 & 4.5 & CSS\\
7C2867         & 0.600 & $-$23.86 & 0.230 & 0.76  & 381.5 & FRII\\
7C2886         & 0.610 & $-$23.52 & 0.148 & 0.71  & 76.7 & FRII \\
7C2928         & 0.720 & $-$24.05 & 0.256 & 0.77  & 314.0 & FRII \\
7C3066         & 0.600 & $-$24.52 & 0.671 & 0.88  & 312.9 & FRII \\
7C3201         & 0.500 & $-$24.23 & 0.085 & 0.81  & 300.0 & FRII \\
7C3222         & 0.670 & $-$24.34 & 0.270 & 0.64  & 347.4 & FRII \\
7C3450         & 0.590 & $-$23.41 & 0.056 & 0.69  & 7.6 & CSS \\
7C3814        & 0.600 & $-$24.81 & 0.094 & 0.76  & 137.4 & FRII \\
3C380         & 0.691 & $-$25.81 & 37.78 & 0.69  & 27.1\footnote{3C380 appears compact, but its intrinsic size is probably larger than 30 kpc since there
is evidence for superluminal motion in this source \cite{wilkinson}.} & FRII \\
\end{tabular}
\label{table:table1}
\end{minipage}
\end{table*}

The eight sources in the high-luminosity sample
were drawn from the Molonglo/APM
Quasar Survey (MAQS) (Serjeant 1996; Maddox et al., in preparation; Serjeant et al., in 
preparation)
which consists of sources with radio flux densities 
$S_{\rm 408MHz}\geq0.95$ Jy from the Molonglo Reference Catalogue (MRC)
\cite{large}
identified with objects on the UK Schmidt plates having
$b_{J}<22.5$.
The MAQS thus have a faint optical completeness limit and a relatively
bright radio flux density limit, and its completeness was estimated
to be 99 per cent by Willott et al.\ \shortcite{willott}.

We also included two more quasars in the sample, 3C380  
and 5C6.189. The most powerful source, 3C380, was selected 
from the 3CRR sample of Laing, Riley \& Longair \shortcite{laing}, which has a flux limit of 
$S_{\rm 178MHz}>10.9$ Jy ($\approx 6.05$ Jy at 408 MHz).
The radio quasar 5C6.189 was selected from the 
$0.1<S_{\rm 151MHz}<0.5$ Jy complete sample of 
Rossitter \shortcite{rossitter}, which has essentially the same selection 
criteria as the 7C sources, so 5C6.189 is included in the 
low-luminosity sample. This quasar has a spectral index of 0.36,
so strictly speaking it does not belong with the other steep-spectrum
sources in the sample. It is unresolved to a limit of 3 arcsec in radio images, and
since it also has a low radio luminosity we believe that it might be the
Doppler boosted core of an intrinsically much less luminous radio source. We
therefore neglected this object in the correlation analysis in Section~\ref{section:section7}.

The majority of the sources in the sample exhibit FRII morphology
with bright cores, jets, edge-brightened lobes and hotspots, typical
of radio-loud quasars. Two of the 7C and
four of the MRC quasars are compact steep-spectrum (CSS) sources. CSS sources have
apparent sub-galactic sizes ($<30$ kpc) in radio and are thought to be
intrinsically compact rather than being extended sources
shortened by projection effects \cite{fanti}.
It is believed that the CSS sources are either young radio sources
in the stage of formation, or that they are confined by
a high-density ISM in the host galaxy which is likely to be denser
than the IGM/ICM confining the more extended radio sources.
It is uncertain whether this will bias the CSS sources in the sample
toward richer or poorer galaxy environments, since it is difficult to see a reason for
a correlation between the local gas density in the core of the host galaxy
and the gas density (or the galaxy density) on Mpc scale in the surrounding
cluster.

In Fig.~\ref{fig:figure1} we have plotted the radio luminosity of the quasars
in the sample as a function of redshift, and various properties 
of the sources are listed in Table~\ref{table:table1}. 
The quasar $B$ luminosities in this table were calculated as 
follows: For the MRC quasars, we obtained 
$m_{B}$ from APM magnitudes assuming $m_{\rm APM}\approx 29.42-b_{J}$ 
(Maddox, personal communication) and $b_{J}\approx m_{B}-0.09$ \cite{metcalfe}.
For the 7C sources, we calculated $m_{B}$ from $E$ magnitudes
by first assuming $R\approx E$ and thereafter converting to $B$ by assuming a power law spectrum
$S_{\nu}\propto \nu^{-0.5}$ and using the $B$ and $R$ zero points given by
Longair \shortcite{longair81}. The $m_{B}$ magnitudes of 5C6.189 and 
3C380 were found from $R$ and $B$ magnitudes given by Rossitter \shortcite{rossitter}
and Netzer et al.\ \shortcite{netzer}, respectively. 
The apparent magnitudes were thereafter converted to 
absolute $M_{B}$ luminosities by assuming a $K$-correction equal to 
$2.5\left(\alpha-1\right)\log\left(1+z\right)$. 

All sources, except 3C380 and 5C6.189, have galactic latitudes $|b|>42\degr$ and galactic 
reddening $E\left(B-V\right)<0.04$. 3C380 and 5C6.189 lie at 
latitudes 23\fdg5 and 27\fdg3 and have galactic reddening of 0.07 and 
0.06, respectively.

\begin{table*}
\begin{minipage}{15cm}
\caption{Positions and details about the observations of the quasar (upper part) and 
background fields (lower part). `NOT' is the Nordic Optical 
Telescope, `McD' is the 107'' telescope at the McDonald Observatory and `HST' is 
the Hubble Space Telescope.}
\begin{tabular}{lllllll}

Source & RA (1950) & DEC (1950) & Filter & Exp. time  & Telesc., date \\  
       &           &            &        & (s)  &          \\
       &           &            &        &           &          \\
MRC 0032$-$203 & 00:32:38.61 & $-$20:20:29.58 & $R$      & 4$\times$600 & NOT 94Dec23 \\
MRC 0033$-$000 & 00:33:53.28 & $-$00:03:27.35 & $R$      & 4$\times$600 & NOT 94Dec25  \\
MRC 0106$-$233 & 01:06:37.57 & $-$23:23:27.63 & $R$, $I$ & 4$\times$600 & NOT 94Dec23,24 \\
MRC 0144$-$058 & 01:44:14.29 & $-$05:52:56.58 & $R$      & 4$\times$600 & NOT 94Dec 24 \\
MRC 0159$-$117 & 01:59:30.39 & $-$11:47:00.12 & $R$      & 4$\times$600 & NOT 94Dec 24 \\
5C6.189        & 02:15:45.30 & $+$31:35:39.60 & $R$      & 4$\times$600 & NOT 94Dec 25 \\
                        &         &      & $R$      & 1$\times$300\footnote{Short exposure
to enable deeper non-photometric images of this field to be properly calibrated.} & NOT 96July21\\
MRC 0222$-$008 & 02:22:34.59 & $-$00:49:03.89  & $R$, $I$ & 4$\times$600 & NOT 94Dec23 \\
MRC 0405$-$123 & 04:05:27.50 & $-$12:19:32.52 & $V$, $R$ & 4$\times$600 & NOT 94Dec23,24 \\
MRC 0406$-$180 & 04:06:52.14 & $-$18:05:02.39  & $R$, $I$ & 4$\times$600 & NOT 94Dec24  \\
7C2671       & 10:19:33.03 & $+$45:56:16.10 & $R$, $I$ & 4$\times$600 & NOT 97May11,10 \\
7C2676       & 10:19:42.68 & $+$39:46:42.99 & F675W    & 4$\times$500 & HST 95Oct16 \\
7C2704       & 10:20:08.78 & $+$48:07:10.41  & F675W    & 4$\times$500 & HST 96Jan12 \\
7C2867       & 10:24:04.46 & $+$41:54:38.75 & F675W    & 4$\times$500 & HST 95Nov15 \\
7C2886       & 10:24:28.39 & $+$43:06:59.89 & F675W    & 4$\times$500 & HST 95Oct15  \\
7C2928       & 10:25:34.68 & $+$43:21:46.30 & $I$      & 9$\times$300 & NOT 97May14  \\
7C3066       & 10:28:47.86 & $+$42:09:46.30 & $V$, $R$ & 4$\times$600 & NOT 94Dec24 \\
7C3201       & 10:32:10.50 & $+$39:26:01.66  & F675W    & 4$\times$500 & HST 95Nov6  \\
7C3222       & 10:32:21.15 & $+$34:21:58.10 & $R$, $I$ & 11$\times$300 & McD 96Feb20 \\
7C3450       & 10:37:26.01 & $+$45:05:14.10& $R$ & 11$\times$300 & McD 96Feb21 \\
7C3814       & 10:45:16.82 & $+$38:53:27.60& $V$, $R$ & 4$\times$600 & NOT 94Dec23\\
3C380        & 18:28:13.55 & $+$48:42:40.38 & $I$      & 8$\times$300 & NOT 97May13,14 \\
\hline
0332$-$1152 & 03:32:45.60 & $-$11:52:07.10 & $I$      & 5$\times$600  & NOT 94Dec23 \\
0405$-$1209 & 04:05:27.43 & $-$12:09:29.10& $R$      & 4$\times$ 600 & NOT 94Dec24 \\
1018$+$4556 & 10:18:58.00 & $+$45:56:22.80 & $I$      & 9$\times$300  & NOT 97May11,13 \\
1025$+$4311 & 10:25:34.43 & $+$43:11:49.00 & $I$      & 9$\times$300  & NOT 97May14 \\
1032$+$3416 & 10:32:21.15 & $+$34:16:58.10 & $R$, $I$ & 11$\times$300 & McD 96Feb20 \\
1037$+$4510 & 10:37:26.01 & $+$45:10:14.10 & $R$      & 6$\times$300  & McD 96Feb21 \\
1217$+$3650 & 12:17:40.13 & $+$36:50:45.50 & $I$      & 11$\times$300 & McD 96Feb18 \\
1349$+$6434 & 13:49:13.18 & $+$64:34:24.00& $I$      & 4$\times$600  & NOT 97May10 \\
1545$+$4745 & 15:45:00.00 & $+$47:45:00.00 & $R$, $I$ & 4$\times$600  & NOT 96July21\\
1755$+$6820 & 17:55:55.07 & $+$68:20:54.10 & $R$, $I$ & 4$\times$600  & NOT 97May11,13,10 \\
1758$+$6500 & 17:58:00.00 & $+$65:00:00.00 & $I$      & 4$\times$600  & NOT 96July24 \\
1807$+$6821 & 18:07:00.05 & $+$68:21:11.50 & $R$      & 4$\times$600  & NOT 97May14 \\
2235$+$0130 & 22:35:00.00 & $+$01:30:00.00 & $V$, $R$ & 4$\times$600  & NOT 96July21 \\
2238$+$0200 & 22:38:00.00 & $+$02:00:00.00 & $R$, $I$ & 4$\times$600  & NOT 96July22,23 \\
2355$+$0240 & 23:55:00.00 & $+$02:40:00.00 & $V$, $R$ & 4$\times$600  & NOT 96July24 \\
\end{tabular}
\label{table:table2}
\end{minipage}
\end{table*}

%*************************************************************

\section{Observations}
\label{section:section4}

Most of the quasar fields were imaged 
with the HiRAC camera at the 2.56-m Nordic Optical Telescope (NOT)
at La~Palma, Spain, during 1994 Dec 23--25
and 1997 May 10--15. The 7C3222 and 7C3450 fields were 
imaged with the 107'' telescope at the McDonald Observatory in Texas
during 1996 Feb 17--22.
Images were also taken of several background control fields to
provide a measurement of the field galaxy counts. The least biased
way of obtaining background galaxy counts is to estimate it from the
edges of the quasar fields, but the HiRAC CCD at the NOT has too small a field
of view (3$\times$3 arcmin) for this approach.
The background images were therefore obtained by offsetting the
telescope 5 or 10 arcmin either north or south of
sources in our AGN sample, or by selecting a random position in the sky at a 
similar galactic latitude as the programme targets.

To extend our low radio luminosity sample,
we used data on five 7C quasars obtained with the Hubble Space
Telescope (Serjeant, Rawlings \& Lacy 1997; Serjeant et al.\ 1999, in preparation). 
These images were taken in the
F675W-filter in 4$\times$500 s exposures with the WFPC2. 
In Table~\ref{table:table2}, positions and details about the observations
of the quasar and background fields are provided.

\begin{table*}
\begin{minipage}{17cm}
\caption{The observing runs.}
\begin{tabular}{llllllll}
Date  & Telesc.  & CCD & FOV          & Pixel scale  & Average & Extinction & Remarks \\ 
      &            &     & (arcmin$^2$) & (arcsec/pix)& seeing (arcsec)      & in $V$ (mag) &       \\
      &            &      &              &                                    &             &     \\   
1994 Dec 23,24\footnote{Most of the radio-loud quasar fields were obtained during Dec 23--24.} & NOT  & 1k SiTe & 3$\times$3    & 0.176 & 0.6 & 0.13 & photometric \\
1994 Dec 25      & NOT        & 1k SiTe & 3$\times$3    & 0.176 & 1.0      & 0.13    & variable transparency\\ 
1996 Feb 17--22  & McD       & 1k TEK   & 11$\times$11  & 0.700 & 2--3     & 0.17    & partly cloudy \\
1996 July 21--23 & NOT        & 1k SiTe & 3$\times$3    & 0.176 & 0.7      & 0.1--0.2& photometric\\
1996 July 24     & NOT        & 1k SiTe & 3$\times$3    & 0.176 & 0.7      & 0.34    & dust in the air\\
1997 May  10\footnote{On May 10 the CCD had lost its quantum efficiency in blue resulting in 
a 0.5 mag reduction in the $I$-band sensitivity.}--14 & NOT  & 2k Loral & 3.7$\times$3.7& 0.110 & 0.7--0.8 & 0.13    & some nights clear,\\
              &             &             &              &        &        &          & some with cirrus\\
\end{tabular}
\label{table:table3}
\end{minipage}
\end{table*}

Eight of the quasar fields were imaged in two filters,
either $V$ and $R$, or $R$ and $I$, depending on the
redshift of the quasar, in order to straddle the 4000 {\AA} break in the 
quasar rest frame.
This was done with the aim of using colours to study the evolution of the 
cluster galaxies, and to 
isolate cluster members with strong 4000 {\AA} breaks. These studies will
be discussed in a future paper.
Quasars with redshifts $z<0.67$  were imaged in $V$ and $R$, whereas
quasars with $z \geq 0.67$ were imaged in $R$ and $I$.
For the fields that were imaged in only one filter, $R$ was chosen
for $z<0.67$ and $I$ for $z\geq0.67$.
Six of the background fields were also imaged in two filters.
With a few exceptions, the exposure time in every quasar and background field is 
4$\times$600 s
(or 9$\times$300 s), so that all fields have a similar depth.
The images from the 107" telescope have exposure 
times of 
11$\times$300 s since the CCD was less sensitive than the CCD at the NOT,
and the seeing was poorer.
The weather and seeing conditions during the observations, as well as 
details about the CCDs that were used, are given in Table~\ref{table:table3}.
For standard star calibration, we observed fields from Christian et al.\ \shortcite{christian} 
during all observing runs.

%*****************************************************************************

\section{Data reduction and photometry}
\label{section:section5}

The preprocessing of the images was performed with 
{\sc iraf}.
Twilight flats were used to flatfield
the NOT data, but the $I$-band data from the 1997 May run were also 
flat--fielded with a skyflat formed by medianing 9--10 frames 
in order to remove fringing.
After co--aligning the images, they were averaged together
and cosmic rays were removed.

The McDonald data were corrected for dark current and flat--fielded with either
twilight or dome flats. Since these data 
suffer from a partially cloudy sky, some extra 
reduction steps were found necessary. All images were divided with a 
skyflat in order to 
remove flat field variations during the night, and to improve the removal 
of the vignetting at the edge 
of the field. To take out the effect of clouds, some of the images 
were scaled so that some reference star in each image had the same flux
as measured in the best image. Each scaled frame was then 
multiplied with a weight--factor inversely proportional to the variance
before co--aligning and combining.

The HST data were corrected for bad pixels using the 
bad pixel maps, and thereafter combined using the {\sc iraf/stsdas}
task {\sc crrej} in order to remove cosmic rays.

\subsection{Photometry and object detection}

For object detection and photometry of faint galaxies in the CCD images
we utilized {\sc focas} (Faint Object Classification and Analysis System, 
e.g.\ Valdes 1989).
The detection limit in {\sc focas} was set to 2.5$\sigma$ above sky level and the `built-in' 
detection filter was used, a 5$\times$5 pixel array designed to increase the sensitivity 
to extended objects with low surface brightness.
Furthermore, objects were detected only if they had sizes larger than
a minimum size which was determined for each frame by measuring the
average seeing from stars in the image. Objects were therefore recognized only
if they had $N$ contiguous pixels above the sky level, where $N$
was varied to match the seeing and the pixel scale of the CCD.
To estimate the magnitudes of the detected objects we used {\sc focas} 
total magnitudes. The 
total magnitudes in {\sc focas} are evaluated by filling
in concavities in the isophote shapes until the area is twice the
isophotal area and then measuring the flux above sky level within this expanded region.
The {\sc focas} total magnitudes have been shown to give good estimates of 
true magnitudes \cite{koo}.

No attempt was made to separate stars and galaxies in the images,
because the wide range in seeing conditions in which the data were taken made it hard to 
perform such separation reliably.
Since the background fields were taken at similar galactic latitudes as the
quasar fields, the number of stars is approximately the same in both. 
Stars will therefore statistically be accounted for in the analysis.
Also, at high galactic latitudes, galaxies are expected to dominate the counts
fainter than $R\sim20$. 
Each detected object was classified by
{\sc focas} by using a template point spread function, which was formed from typically
4--5 selected stars in each image. When we counted the number
of objects in the images, we excluded objects that were 
classified as `n' (noise) or `long', and also in a few cases, objects 
coinciding with the spikes of bright stars.

Magnitudes measured in the F675W-filter were calibrated to Johnson $R$ by using
photometry of objects in fields that were imaged both with 
the HST and the NOT. 
With this approach, we found a zero point of 21.78$\pm$0.47 mag, which is similar to
the one obtained by using the {\em photflam} and {\em photzpt} header keywords to compute
a magnitude in the STMAG system, and then convert to Johnson $R$ as described in the WFPC2 
Instrument Handbook \cite{biretta}.

All magnitudes were corrected for galactic extinction using the maps of 
Burstein \& Heiles \shortcite{burstein} and the Galactic extinction law by
Cardelli, Clayton \& Mathis \shortcite{cardelli}.

\subsection{Completeness simulations}
\label{sec:completeness}

To find the completeness limit 
and to gain as much information 
at faint magnitudes as possible, we simulated
the loss of galaxies (in the ground-based data) in both the quasar
and the background images due to incompleteness.
We utilized {\sc artdata} in {\sc iraf} to add a
grid of $\sim$70 artificial galaxies with known magnitudes 
(typically 20--25 mag) and half--light radii distribution according to
the HST Medium Deep Survey \cite{schmidtke}. 
Two different galaxy profiles were chosen in order to investigate
which galaxies were most easily picked up by {\sc focas},
exponential discs resembling spiral galaxies and de Vaucouleurs profiles
resembling elliptical galaxies. Inclination and position angles
were randomly distributed.

After adding the artificial galaxies, the CCD frames were processed in
{\sc focas} and the result of the detection process examined to find how
many of the added galaxies were detected, and what their total magnitudes had
been estimated to.
This process was repeated several times, each time turning the flux of the
galaxies down by 0.1 mag, until the fraction picked up by {\sc focas}
had decreased significantly.

By comparing the {\sc focas}-computed total magnitudes with the magnitudes originally
assigned to the galaxies, we were able to estimate the errors made by
{\sc focas} as a function of magnitude.
These simulations show that the NOT data are complete down to 23.5
in $R$ and 23.0 mag in $I$ with errors of $\pm$0.3 mag at the faintest limit. The data
from the 107'' telescope are less deep with
a completeness limit of 22.5 in $R$ and 21.5 mag in $I$ with errors of $\pm$0.4
mag at the limit.
Furthermore, galaxies with exponential disc profiles were more easily
detected than those with de Vaucouleurs profiles.

In intervals of half a magnitude, we calculated the fraction of galaxies
detected in {\sc focas} by
averaging the detected fractions of exponential and
de Vaucouleurs profiles.
In order to correct for incompleteness at faint magnitudes, the number of 
galaxies in every magnitude bin 
was divided by the detected fraction in the corresponding bin, but only as 
long as the correction was less than a factor of 2.
The corrected counts were used to find deeper and incompleteness-corrected 
values of the correlation amplitudes.
The WFPC2 images are deeper than the
ground-based images in the sense that fainter galaxies will
be more easily detected. However, since we only have ground-based 
background frames to normalize the counts, we took the 
same magnitude cut--off in the HST images as in the NOT images
when computing \agq and $B_{\rm gq}$.

%********************************************************************

\section{Results}
\label{section:section6}

\subsection{Field galaxy counts}

The background control images obtained at the NOT yield an area of 
18 arcmin$^2$ in $V$, 58.7 arcmin$^2$ in $R$ and 73.7 arcmin$^2$
in $I$, most of them taken during the observations in July 1996 and May 1997.
The background images from the 107'' telescope cover an area of 103.3 
and 97.7 arcmin$^2$ in $R$ and $I$, respectively.
Fig.~\ref{fig:figure2} and \ref{fig:figure3} show the number counts per deg$^2$ of 
field galaxies as obtained from the NOT images. The average number counts of galaxies in the
quasar fields are also plotted. The number counts in these two figures have been corrected
for incompleteness as described in section \ref{sec:completeness}, and the
corrected field counts have slopes of 0.29$\pm$0.05 and 0.34$\pm$0.06 in $R$ and $I$, 
respectively.
In the $R$-band, the slope of the raw number counts is 0.24$\pm$0.03, whereas 
for the raw $I$ counts, the slope is 0.26$\pm$0.06.
Our estimates of the slopes are in agreement with that
found by others (Smail et al.\ 1995; Steidel \& Hamilton 1993; Lilly, Cowie \& Gardner 1991).
For example, \ Smail et al. find a slope of 0.321$\pm$0.001 
in $R$ and 0.271$\pm$0.009 in $I$, which is within the errors of the slopes of 
our completeness corrected counts.

The error bars in Fig.~\ref{fig:figure2} and \ref{fig:figure3} were calculated as 
Poisson errors multiplied with a factor ($\approx$1.3) that approximates the departure
from Poisson statistics if there is a non-random distribution of
field galaxies \cite{yeelopez}.

\subsection{Calculation of \agq and \bgq}

In the quasar fields, we counted galaxies with magnitudes brighter than the
completeness limit within a 
circle of 0.5 Mpc centered on the quasar (0.5
Mpc at $z=0.7$ corresponds to $\approx66$ arcsec). The average background 
galaxy count
down to the same magnitude was thereafter 
subtracted in order to give the excess number of 
galaxies, $N_{\rm tot}-N_{\rm b}$.
Typical values of $N_{\rm tot}$ in the richest fields were 
70--80 whereas the poorest fields had $N_{\rm tot}\approx$ 20--30. The
background count, $N_{\rm b}$, was typically around 45.
A corrected excess was also calculated using the incompleteness corrected counts. 
In most fields this could be done down to a fainter magnitude
than what we defined to be the completeness limit, but not fainter than 
24.5 in $R$ and 23.5 mag in $I$, since here the correction factor was 
greater than 2. 

In order to prevent systematic errors, the quasar and background frames were treated
identically during data reduction and object detection. Also, the
background galaxy counts obtained at one of the two (NOT and 107'')
different telescopes were applied only to quasar frames from the same telescope. 
The galaxy excess in the HST images was found by subtracting
the NOT background counts. For fields that were 
imaged in two filters, we used the galaxy counts in the image taken through
the reddest filter for computing $B_{\rm gq}$, since this gives preference to 
red galaxies that are likely to lie at the quasar redshift.

The field size of 3$\times$3 arcmin at the NOT was sufficient
to cover the 0.5 Mpc radius, provided the quasar was
well centered on the frame. Two of the quasar frames were nevertheless found
to have parts of the circle lying outside the edges of the field. 
No large areas were 
missing, 0.28 arcmin$^2$ at the most, which is less than 8 per cent of the 
area of the circle. We also took out $\approx$ 6.5 per cent of the area in the
MRC0222$-$008 field because of a large, bright foreground galaxy.
The missed and the chopped-out areas were corrected for by assuming 
a density of galaxies equal to the mean density within the remaining area.
The same approach was applied to the HST data, where the 
`L'-shaped field of view of the WFPC2 covered approximately half the 
area that we were interested in.

\begin{figure}
\psfig{file=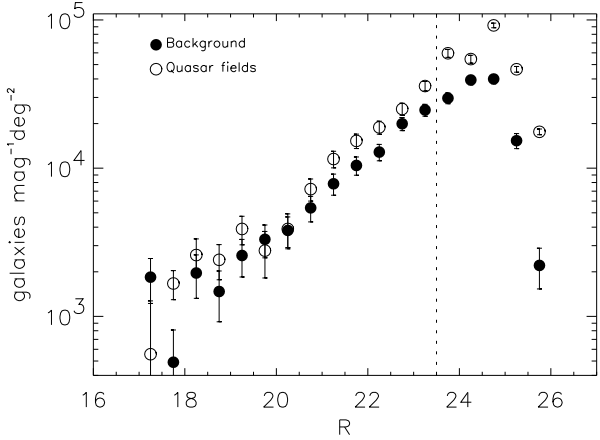}
\caption{This plot compares the average number count of galaxies in the 
quasar (open circles) and the background images (filled circles). A clear excess
of galaxies in the quasar fields is seen from $R\approx20$. The completeness limit is 
indicated by the vertical dotted line. The slope of the background counts at 
$R>20$ (where the counts are not contaminated by stars) was found to be 0.29$\pm$0.05.}
\label{fig:figure2}
\end{figure}

\begin{figure}
\psfig{file=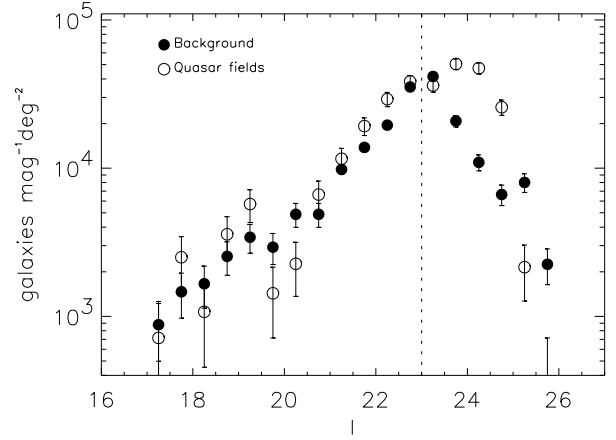}
\caption{Same as in Fig.~2, but for counts in the $I$-band. The slope
of the background counts at $I>20$ is 0.34$\pm$0.06. The `bump' at $I<20$ is probably
caused by stars.}
\label{fig:figure3}
\end{figure}

The excess number of galaxies in each field was used to find the angular 
cross-correlation amplitude, $A_{\rm gq}$. Computed values of $A_{\rm gq}$, 
both from the raw counts and from the corrected counts, are shown in 
Table~\ref{table:table4}. 
We also averaged together the galaxy number counts within the 
0.5 Mpc radius in all the quasar fields and plotted it together with
the field galaxy counts as shown in Fig.~\ref{fig:figure2} and \ref{fig:figure3}
to demonstrate that there is a real excess of galaxies.

In order to obtain the spatial cross-correlation amplitude, $B_{\rm gq}$, 
the angular correlation amplitude has to be normalized with the number of
field galaxies at the quasar redshift, as given by the integrated luminosity
function, $\Phi\left(m_{\rm lim},z\right)$.
We used the Schechter function \cite{schechter} with parameters 
$\alpha=-0.89$ and $M_{AB}^{*}\left(B\right)=-20.83$ based on Lilly 
et al.'s \shortcite{lilly} estimation of the CFRS luminosity function
at redshifts $0.5\leq z\leq0.75$. The values of 
$\phi^{*}$ employed were $0.0072$ and 0.0052 Mpc$^{-3}$ in $R$ and $I$, respectively.
Below, we explain how the $\phi^{*}$'s were chosen in order to be consistent with
the observed field galaxy counts.
The CFRS luminosity function
is well suited to our analysis, especially since CFRS is an $I$-band selected
survey, and results from Lin et al.\ \shortcite{lin} 
based on the CNOC2 (Canadian Network for Observational Cosmology)
Field Galaxy Redshift Survey show that there is general agreement between
the CFRS and the CNOC2 luminosity function.

To make sure that the observed background galaxy counts were consistent with the $\phi^{*}$ 
normalization, we constructed galaxy counts from a combination of 
four Schechter function segments. These constructed counts were 
compared to the observed counts to find the $\phi^{*}$ that best matched the data.
We selected luminosity function segments for four
redshift intervals, $\left<0.0,0.2\right>$, $\left<0.2,0.5\right>$, 
$\left<0.5,0.75\right>$ and $\left<0.75,1.0\right>$.
In the first interval, we used the Schechter parameters found by 
Loveday et al. \shortcite{loveday} for the local luminosity 
function in the Stromlo-APM survey; $M_{b_{J}}=-19.5$, $\alpha=-0.97$ and 
$\phi^{*}=0.014$ Mpc$^{-3}$. For the three other intervals, we used 
$M^{*}_{AB}\left(B\right)=-21.04$, $-20.83$ and $-21.24$ as found by 
Lilly et al. for their `All' colour sample in the CFRS. 
We also fixed $\alpha$ at $-$0.89 in these three redshift intervals, since the slopes
reported by Lilly et al.\ at these redshifts are seen to vary a lot, which presumably is
not a real evolutionary effect. Since the CFRS luminosity function is defined
over a narrow magnitude range, the faint-end slope may have high uncertainties.
For example, \ in the redshift range $\left<0.5,0.75\right>$ that we are most interested in here,
and at luminosities from $M_{AB}\left(B\right)=-19.7$ to $-$22.9, Lilly et al.\ find that 
$\alpha=-0.5$. The faintest absolute magnitude in this range corresponds to 
$\approx-21$ in $R$, whereas our 
completeness limit in $R$ corresponds to $\approx-20$ at $z=0.6$, so our data are approximately one
magnitude deeper. We therefore chose $\alpha=-0.89$ since this is the slope 
they find for the total luminosity function from $z=0.0$ to 1.3.

The four luminosity function segments were integrated with respect to luminosity 
down to the completeness limit of our data at the average redshift in each interval. 
The completeness magnitudes in $R$ and $I$ were found
by applying $K$-corrections and galaxy colours of an Sa galaxy listed as a function of 
redshift by Rocca--Volmerange \& Guiderdoni \shortcite{rocca-volmerange} and Guiderdoni \& 
Rocca--Volmerange \shortcite{guiderdoni}.
The characteristic magnitudes in each interval were transformed from $B_{AB}$ to $R$ and $I$ 
by first using that $B_{AB} = B-0.17$ \cite{oke1}, and thereafter assuming 
$B-V=0.74$, $V-R=0.68$ and $R-I=0.57$ which are the colours of an Sa galaxy model 
at $z=0$ \cite{guiderdoni}. The $M^{*}_{b_{J}}$ magnitude in the lowest redshift interval
was converted to $B$ by assuming $b_{J}=B-0.09$ \cite{metcalfe}.

Then the integration was carried out over redshift, from $z=0.0$ to 1.0. From these
approximate calculations we constructed number counts that were subsequently matched to 
the observed counts by varying the $\phi^{*}$'s of each segment until a minimum in 
$\chi^{2}$ was reached.
For the $I$-band data we found it sufficient to use the same scaling for all
intervals, and the best fit was obtained with $\phi^{*}=0.0052$ Mpc$^{-3}$ for 
the $\left<0.5,0.75\right>$ interval.
In $R$, very little adjustment was needed in $\phi^{*}$ to fit the data, and for 
the $\left<0.5,0.75\right>$ interval we found a best fit with 
$\phi^{*}=0.0072$ Mpc$^{-3}$.

To normalize $B_{\rm gq}$, we evaluated  
$\Phi\left(m_{\rm lim},z\right)$ for every quasar field by  
integrating the Schechter function from $-\infty$ to an absolute magnitude 
corresponding to the completeness limit at the quasar redshift.
The absolute limiting magnitude in every quasar frame was 
found by converting the apparent completeness limit
to an absolute magnitude using
$K$-corrections for a hot E/S0 galaxy \cite{rocca-volmerange}.
Typical values obtained for the integrated luminosity function were
$\Phi\left(m_{\rm lim},z\right)\sim0.005$--$0.007$ Mpc$^{-3}$.

\begin{table*}
\begin{minipage}{14cm}
\caption{Clustering statistics for the quasar fields consisting of the angular and the spatial
galaxy--quasar cross-correlation amplitudes, \agq and $B_{\rm gq}$, and the more 
direct Abell-type measurement, $N_{0.5}$.
The correlation amplitudes were calculated both from raw galaxy counts and from galaxy counts 
corrected for incompleteness. The magnitude limits were set to $R=23.5$
and $I=23.0$ for the raw counts, whereas a deeper limit was applied for the corrected counts, 
maximum 24.5 in $R$ and 23.5 in $I$.
The fields that are denoted `unchanged' are fields that could not be pushed above the 
completeness limits of 23.5 in $R$ and 23.0 mag in $I$. The corrected counts
were not applied to the HST images either, as denoted by `...'. The errors in \agq and \bgq
are $\Delta A_{\rm gq}$ and $\Delta B_{\rm gq}$ from Eq.~3.}
\begin{tabular}{llrrrrrr}
Source                  & $z$   & $A_{\rm gq}\times10^{-3}$   & $A_{\rm gq}\times10^{-3}$   & \bgq            & \bgq            & \n \\
                        &       & (rad$^{0.77}$)        &  (corrected)          & (Mpc$^{1.77}$)  &   (corrected)   &    \\
                        &       &                        &              &               &            &   \\
MRC 0032$-$203 & 0.518 & 0.52$\pm$0.26    & 0.48$\pm$0.20    & 253$\pm$127     & 303$\pm$129     & 6.0$\pm$8.2 \\
MRC 0033$-$000 & 0.560 & $-$0.24$\pm$0.22 & $-$0.13$\pm$0.18 & $-$142$\pm$133  & $-$100$\pm$134  & $-$3.5$\pm$7.8 \\
MRC 0106$-$233 & 0.818 & 0.00$\pm$0.25    & 0.30$\pm$0.21    & $-$3$\pm$233    & 345$\pm$236     & $-$3.8$\pm$7.7 \\
MRC 0144$-$058 & 0.630 & 0.62$\pm$0.27    & 0.66$\pm$0.19    & 421$\pm$184     & 539$\pm$157     & 25.0$\pm$9.7 \\
MRC 0159$-$117 & 0.669 & 0.31$\pm$0.26    & 0.54$\pm$0.19    & 272$\pm$222     & 507$\pm$177     & 5.6$\pm$8.7 \\
5C6.189        & 0.597 & 1.07$\pm$0.29    & 0.75$\pm$0.22    & 685$\pm$186     & 590$\pm$170     & 15.4$\pm$9.0 \\
MRC 0222$-$008 & 0.687 & 0.36$\pm$0.27    & 0.58$\pm$0.22    & 246$\pm$185     & 537$\pm$197     & 2.0$\pm$7.6 \\
MRC 0405$-$123 & 0.574 & 0.94$\pm$0.28    & 1.15$\pm$0.23    & 579$\pm$179     & 875$\pm$176     & 12.7$\pm$8.8 \\
MRC 0406$-$180 & 0.720 & 0.79$\pm$0.29    & 0.50$\pm$0.21    & 569$\pm$174     & 469$\pm$201     & $-$0.6$\pm$7.6 \\
7C2671         & 0.745 & 0.04$\pm$0.25    & unchanged        & 35$\pm$212      & unchanged       & $-$1.9$\pm$7.5 \\
7C2676         & 0.604 & $-$0.41$\pm$0.21 & ...              & $-$265$\pm$138  &  ...            & $-$7.3$\pm$6.8\\
7C2704         & 0.580 & 0.41$\pm$0.26    & ...              & 253$\pm$160     &  ...            & 11.6$\pm$8.0\\
7C2867         & 0.600 & $-$0.04$\pm$0.23 & ...              & $-$25$\pm$152   &   ...           & $-$0.2$\pm$7.3\\
7C2886         & 0.610 & 0.11$\pm$0.24    & ...              & 70$\pm$159      &  ...            & 3.4$\pm$7.6\\
7C2928         & 0.720 & $-$0.13$\pm$0.24 & unchanged        & $-$94$\pm$175   & unchanged       & $-$2.6$\pm$7.4 \\
7C3066         & 0.600 & 0.22$\pm$0.25    & $-$0.21$\pm$0.16 & 144$\pm$161     & $-$168$\pm$126   & 0.4$\pm$8.1 \\
7C3201         & 0.500 & $-$0.06$\pm$0.23 &     ...          & $-$29$\pm$109      &     ...         & 5.6$\pm$7.6\\
7C3222         & 0.670 & $-$0.05$\pm$0.38 & 0.03$\pm$0.33    & $-$52$\pm$364   & 27$\pm$259     & $-$0.1$\pm$5.7\footnote{Counted
within $m_{\rm g}$ and $m_{\rm g}+2.27$}\\
7C3450         & 0.590 & 1.33$\pm$0.44    & 1.49$\pm$0.38    & 910$\pm$297     & 895$\pm$227     & 26.4$\pm$7.7\footnote{Counted
within $m_{\rm g}$ and $m_{\rm g}+2.66$}\\
7C3814         & 0.600 & 0.28$\pm$0.25    & 0.44$\pm$0.21    & 178$\pm$163      & 350$\pm$162     & 17.4$\pm$9.1 \\ 
3C380          & 0.691 & 0.66$\pm$0.28    & unchanged        & 458$\pm$198      & unchanged       & 7.0$\pm$7.9 \\
\end{tabular}
\label{table:table4}
\end{minipage}
\end{table*}

The results of the computations of \bgq are shown in Table~\ref{table:table4}.
The errors in \agq and \bgq were calculated according to 
\begin{equation}
\frac{\Delta A_{\rm gq}}{A_{\rm gq}}=\frac{\Delta B_{\rm gq}}{B_{\rm gq}}=\frac{\left[(N_{\rm tot}-N_{\rm b})+1.3^{2}N_{\rm b}\right]^{1/2}}{N_{\rm tot}-N_{\rm b}}
\label{eq:equation3}
\end{equation}
\noindent 
\cite{yeelopez} and reflect the 
errors in the galaxy counts on the quasar and background images, but
not uncertainties in the luminosity function.
(Yee \& L{\'o}pez--Cruz also argue that the inverse of this error underestimates 
the statistical significance of excess galaxy counts which may be better represented by
$(N_{\rm tot}-N_{\rm b})/1.3\sqrt{N_{\rm b}}$).
In Table~\ref{table:table4}, \bgq found from both the raw and the corrected counts
are shown. On average the corrected $B_{\rm gq}$'s are larger since these were deduced from
deeper counts, in some cases down to $R=24.5$
(for MRC0144$-$058, MRC0159$-$117 and 7C3066).
In cases where the quasar and background fields had different completeness limits, we took the
brightest magnitude as the limit. Some fields could not be pushed further with regard to the limiting
magnitude, so these are left as `unchanged'.
The HST images were also not corrected.

The counting of galaxies should not be done either too far below or too far above $M^{*}$ 
of the surrounding galaxy cluster in order to obtain a reliable result. At faint magnitudes 
the background counts rise faster than the associated galaxy counts,  
and a possible excess may drown in the background.
Yee \& L{\'o}pez--Cruz \shortcite{yeelopez} showed that if the appropriate luminosity 
function is used to normalize the counts, i.e.\ that the $\phi^{*}$ of this luminosity 
function can reproduce the observed background 
counts, then the best results are obtained by 
counting down to a magnitude between $M^{*}+1$ and $M^{*}+3$. 
As far as possible, we have attempted to reach $M^{*}+2$, and 
our completeness magnitudes of $R=23.5$ and $I=23.0$ correspond to $M_{R}\approx-20.5$ 
and $M_{I}\approx-20.8$  at redshifts
0.6 and 0.7 (assuming $K$-corrections of approximately 1 and 0.5 mag). 
Provided that $M^{*}$ of the associated galaxies is approximately
the same as the one in the field given by Lilly et al.\ \shortcite{lilly}, we 
have reached deep enough in all fields, except 7C3222 and 7C3450. These two
fields have
shallower completeness limits (imaged at the 107'' telescope) of 22.5 in $R$ and
21.5 in $I$, corresponding to 
$M^{*}_{R}\approx-21.5$ and $M^{*}_{I}\approx-22.4$, i.e.\ not as deep as 
$M^{*}+1$. Using the corrected counts, we go half a magnitude deeper and should therefore be
on the faint side of $M^{*}+1$. 

\subsection{Comparison with previous work}

\begin{figure}
\psfig{file=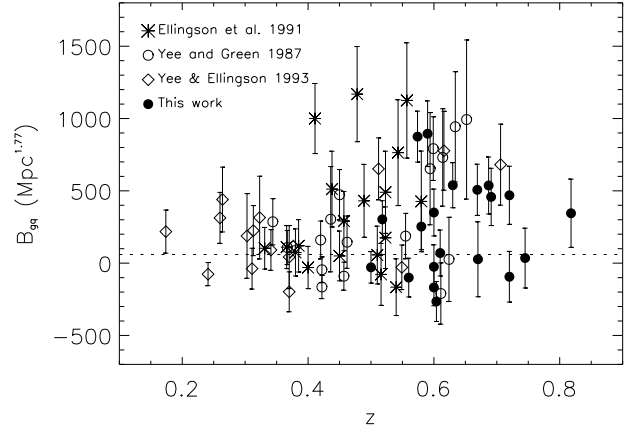}
\caption{The cross-correlation amplitudes from this work, derived from the corrected
counts, plotted together with \bgq for steep spectrum radio-loud quasars from the literature. 
The galaxy--galaxy correlation amplitude for low-$z$ galaxies in the field, 
$B_{\rm gg}\approx60$ Mpc$^{1.77}$ (Davis \& Peebles 1983), is shown as a dotted 
line across the plot.}
\label{fig:figure4}
\end{figure}

We may compare with
YG87 and EYG91 who also quantified quasar environments in terms of $B_{\rm gq}$.
The greatest difference between their and our analysis is the method used for choosing
the appropriate luminosity function for normalizing $B_{\rm gq}$.
Since determinations of the field galaxy luminosity function at higher redshifts
were not available to them, they used 
local luminosity functions from the literature which they fitted to the observations
in two steps. 
First, they added an evolution in $M^{*}$ that matched the observed luminosity
function of the associated galaxies. A possible risk with this is that too much
evolution may be added to $M^{*}$ if the quasar fields are contaminated by foreground
galaxies. 
This approach also assumes
that the field and cluster luminosity function evolve together. 
Second, they scaled $\phi^{*}$ of the evolved local luminosity function to 
match the observed background counts. 

For $\left<z\right>=0.6$, YG87 used $M^{*}_{r}=-21.55$, $\alpha=-1.2$ and $\phi^{*}=0.0058$
Mpc$^{-3}$ ($q_{0}=0.5$), and EYG91 used $M^{*}_{r}=-21.73$, $\alpha=-1.0$ and $\phi^{*}=0.0027$
Mpc$^{-3}$ ($q_{0}=0.02$).
To compare the characteristic magnitudes, we transformed from $r$ to $R$ by 
assuming $r-R\approx0.43$ \cite{fukugita} and applied $K$-corrections for a 
hot E/S0 galaxy \cite{guiderdoni} since YG87 and EYG91 give $M^{*}_{r}$ in the 
observed waveband. 
We found $M_{R}^{*}\left({\rm YG87}\right)=-22.98$ and  
$M_{R}^{*}\left({\rm EYG91}\right)=-22.85$ ($q_{0}=0.5$). By adding a colour of 
$R-I=0.78$ to the $R$ magnitudes, we obtained $M_{I}^{*}\left({\rm YG87}\right)=-23.76$ 
and $M_{I}^{*}\left({\rm EYG91}\right)=-23.63$.
It therefore seems that we have used an $M^{*}$ that is approximately one mag 
fainter than in both of these studies, and also a $\phi^{*}$ that is larger.
A fainter characteristic magnitude will lower the value of the integrated luminosity function, 
thereby bringing \bgq up by a small amount, but at the same time, a larger value of 
$\phi^{*}$ will increase the number of galaxies above the limiting magnitude, 
with the result that \bgq decreases.

The mean \bgq for the quasar fields in our sample is 265 Mpc$^{1.77}$ with an error in
the mean of 74 Mpc$^{1.77}$ (from a combination of intrinsic dispersion in $B_{\rm gq}$
and measurement error). With 5C6.189 excluded, this becomes 249$\pm$75 Mpc$^{1.77}$.
To see how the choice of different luminosity functions affected our measurements, 
we recomputed \bgq with both YG87's and EYG91's
luminosity functions. In the first case, we obtained overall lower values of $B_{\rm gq}$,
the mean was found to be 185$\pm$57 Mpc$^{1.77}$. 
In the second case, we transformed to a $q_{0}=0.02$ cosmology for consistency,
and found a better agreement, the mean \bgq 
was 257$\pm$77 Mpc$^{1.77}$. The correlation amplitude
is rather insensitive to $q_{0}$, and 
using $q_{0}=0.02$ has very little effect on $B_{\rm gq}$. E.g.\ at $z=0.5$, 
\bgq is smaller by a factor of $\sim1.15$ if $q_{0}=0.02$ instead of 0.5.  

We have one field in common with 
Yee \& Ellingson \shortcite{yee1} with which we may compare more directly. They report
that \bgq in the MRC0405$-$123 field is 905$\pm$277, 
whereas we find 875$\pm$176 Mpc$^{1.77}$. When we recalculated \bgq with EYG91's luminosity
function, we found 1006$\pm$231 for this field, and with YG87's luminosity function as normalization,
we obtained 647$\pm$96 Mpc$^{1.77}$. There thus seems to be very good agreement between
our and the Yee \& Ellingson result for this field, and in particular
that this quasar lies in a relatively rich cluster.

In Fig.~\ref{fig:figure4} we have plotted our $B_{\rm gq}$ measurements obtained
from the corrected counts together with \bgq for YG87's and EYG91's quasar fields 
as a function of redshift. In this figure, we have also added 
quasar fields reported by Yee \& Ellingson \shortcite{yee1} that  
do not overlap with sources from YG87 and EYG91 (referenced in table 1 in 
Yee \& Ellingson as coming from Yee \& Green \shortcite{yee84} and from unpublished CTIO and CFHT 
images).
We used photometric information available in the NASA/IPAC Extragalactic Database (NED)
to find the spectral index of the literature sources, and
found that some of them were flat-spectrum quasars. These were omitted 
in Fig.~\ref{fig:figure4}, since we only want to compare with other steep-spectrum sources.
(We also found that three of the quasars, 0208$-$018, 0438$-$165 and 0449$-$183, listed
as radio-loud in table~2 of EYG91, are radio-quiet according to the definition of
Kellerman et al.\ \shortcite{kellerman}. These were also ignored.)
It seems that there are no systematic differences in our estimates of \bgq and the 
other estimates from the literature.
The majority of the literature quasar fields have amplitudes
below 500 Mpc$^{1.77}$, a trend which is present also in our fields, and 
the overall impression from Fig.~\ref{fig:figure4} is that the correlation 
amplitudes from the different samples are consistent with each other and fall 
approximately in the same range. 
We do not see any redshift dependence in 
$B_{\rm gq}$, either in our own sample or in 
the combined sample, as shown in Fig.~\ref{fig:figure4}.
Adding the flat-spectrum quasars seems to make no substantial difference to
the plot, and it is difficult to see any significant trend in the flat-spectrum
population alone.

\subsection{Comparison with other two--point correlation functions}

The average \bgq for our sample, 265$\pm$74 Mpc$^{1.77}$, 
is significantly larger than the galaxy--galaxy correlation amplitude in the field 
measured both locally and at higher redshifts. The local galaxy--galaxy correlation function,
$\xi_{\rm gg}\left(r\right)=\left(r/r_{0}\right)^{-\gamma}$, has been measured
by several investigators (e.g.\ Davis \& Peebles 1983; Loveday et al.\ 1995; Guzzo et al.\ 1997)
and results emerging from these studies show that $r_{0}\approx10$
Mpc and $\gamma\sim1.7$, i.e.\ that $B_{gg}\approx60$ Mpc$^{1.77}$ ($H_{0}=50$ km s$^{-1}$ Mpc$^{-1}$).
Written as a correlation length, our mean \bgq corresponds to $r_{0}=23.4$ Mpc.  

Several attempts have been made to measure $\xi_{\rm gg}$ up to $z\approx0.5$, but
the results so far seem to disagree to some extent. Le~F{\`e}vre et al.\ 
\shortcite{lefevre},
using the CFRS, find evidence for a decrease in the correlation length toward
higher redshifts with $r_{0}=2.66$ Mpc and $\gamma=1.64$ at $z=0.53$.
Hudon \& Lilly \shortcite{hudon} find $r_{0}=3.78$ Mpc
at $z=0.48$. On the other hand, Carlberg et al.\ \shortcite{carlberg} and Small et al.\ \shortcite{small}
find no evidence for evolution between $z\approx0.1$ and 0.6, 
and Small et al.\ get a correlation length of $r_{0}=7.4$ Mpc at $z\sim0.3$. 
Nevertheless, the values we are measuring for \bgq are larger than
the observed galaxy--galaxy correlation amplitude at both higher and lower
redshifts, and show that we have clearly detected enhancements in the 
galaxy density around the quasars.

\begin{figure}
\psfig{file=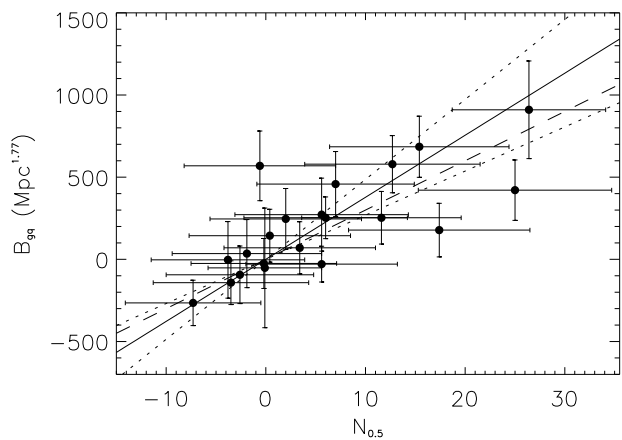}
\caption{\bgq plotted versus $N_{0.5}$. The solid and the dotted lines show the best fit of 
$B_{\rm gq}=\left(37.8\pm10.9\right)N_{0.5}$. The dashed line is the relation
for radio galaxies from Hill \& Lilly (1991), $B_{\rm gg}=30N_{0.5}$.}
\label{fig:figure5}
\end{figure}

\begin{figure}
\psfig{file=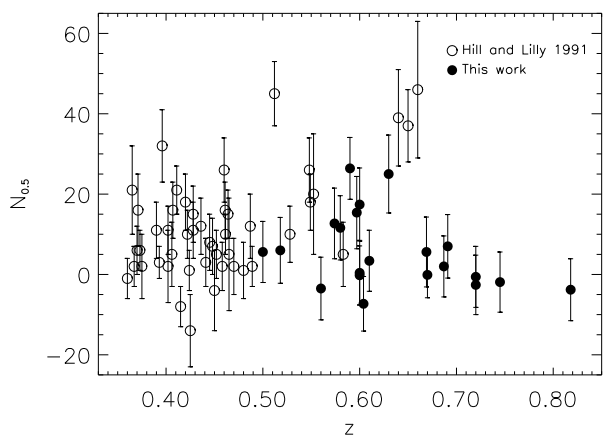}
\caption{\n for the sample of radio galaxies by Hill \& Lilly \protect\shortcite{hill} (open circles)
plotted together with \n for the quasars in our sample (filled circles).}
\label{fig:figure6}
\end{figure}

We can also compare our \bgq values with $B_{\rm cg}$, the amplitude
of the cluster--galaxy
cross-correlation function, $\xi_{\rm cg}$.
Lilje \& Efstathiou \shortcite{lilje} measured $\xi_{\rm cg}$ on scales
down to 0.2 Mpc using Abell clusters 
and the Lick galaxy counts. They found that
$\xi_{\rm cg}$ was well fit by the same functional form as $\xi_{\rm gg}$, 
but with $\gamma=2.2$ and a correlation length $r_{0}=17.6$ Mpc. 
Subsequently, Croft, Dalton \& Efstathiou \shortcite{croft}
have shown that this is also a good approximation to $\xi_{\rm cg}$
obtained from the independent Stromlo-APM galaxy and 
APM cluster redshift surveys on scales of 0.2 to $>20$ Mpc. 
Prestage \& Peacock (1988, 1989), Andersen \& Owen \shortcite{andersen}
and Yee \& L{\'o}pez-Cruz \shortcite{yeelopez} have examined $B_{\rm cg}$ measured in 1, 
1--1.5 and 0.5 Mpc circles, respectively, as a function of cluster richness, 
all assuming $\gamma=1.77$. 
Comparison of these measurements of $B_{\rm cg}$ with the one of Lilje \& Efstathiou
reveals some discrepancies, presumably caused by systematic differences in the 
measurements and scatter in the richnesses of the clusters observed.
To compare the Lilje \& Efstathiou function with the rest, we evaluated
$\xi_{\rm cg}$ at a radius of 0.5 Mpc, then converted to  
$B_{\rm cg}$ assuming $\gamma=1.77$. This comes out to 
$B_{\rm cg}=740$ Mpc$^{1.77}$. Prestage \& Peacock estimate 
$B_{\rm cg}\approx 690$ Mpc$^{1.77}$ for an Abell class 1 cluster, and 
Anderson \& Owen $\approx 615$ Mpc$^{1.77}$, both broadly
consistent with the Lilje \& Efstathiou function (which was calculated for 
clusters with Abell richness class $\geq 1$). In contrast, Yee \& L{\'o}pez-Cruz 
using CCD data find a median $B_{\rm cg}\approx 950$ Mpc$^{1.77}$ for an 
Abell class 1 cluster. We have decided to take the mean for Abell class 1 to be 
740 Mpc$^{1.77}$ simply on the basis that more clusters were involved in 
the Lilje \& Efstathiou study, and note that this is more likely to be an 
overestimate than an underestimate as clusters richer than Abell class 1 
were included. We have two quasar fields that have \bgq larger than this value,
which suggests that a few quasars at $0.5<z<0.8$ are found in clusters of approximately 
Abell class 1 or greater, but that most are in poorer groups or clusters.

We believe that we have not overestimated the richness, since we 
also find fields with negative $B_{\rm gq}$. However, a small or negative \bgq
need not mean that there are no galaxies associated with the quasar, but simply that
the associated galaxies are lost in the fluctuations of the background counts.

\subsection{Calculation of \n}

We also calculated \n \cite{hill} for each quasar 
field by counting the number of galaxies within the 0.5 Mpc radius 
with magnitudes ranging from $m_{\rm g}$ to $m_{\rm g}+3$, 
where $m_{\rm g}$ is the typical
magnitude of a radio galaxy (Eq.~\ref{eq:equation2}) at the quasar redshift.
The average number of galaxies within the same magnitude range in the background
control images was subtracted from the counts in the quasar fields.
The estimates of \n obtained in this manner 
are listed in Table~\ref{table:table4} in the last column.
Since the 7C3222 and 7C3450 fields from the 107'' telescope at the McDonald
Observatory are not deep 
enough for $m_{\rm g}+3$ to be brighter than the completeness limit, we 
have
calculated \n from $m_{\rm g}$ down to the completeness limit (21.5 in $I$ 
and 22.5 in $R$).

Although this is a measurement intended for fields around 
radio galaxies, it seems to give a good estimate of the richness in quasar fields  
since we find an overall correspondence between  \bgq and 
$N_{\rm 0.5}$, as shown in Fig.~\ref{fig:figure5}. In this figure, large values of \bgq 
is seen to pair rather well with large numbers of $N_{0.5}$, and fitting a straight line
gives a best fit of $B_{\rm gq}=\left(37.8\pm10.9\right)N_{0.5}$,
slightly steeper than the relation, $B_{\rm gg}=30N_{0.5}$, found by 
Hill \& Lilly for radio galaxies, but still a good agreement
considering that we have only estimated $m_{\rm g}$ from a magnitude--redshift relation.
There are a couple of fields with higher \n
that do not follow the trend as nicely as the rest of the fields
(7C3814 and MRC 0144$-$058). Some fluctuations are expected,
as \n is not as deep a measurement as $B_{\rm gq}$, and also the 
number of field galaxies in the quasar images fluctuates to some level.

In Fig.~\ref{fig:figure6}, we plot \n for quasars in our sample
together with \n obtained for radio galaxies by Hill \& Lilly \shortcite{hill}.
(Hill \& Lilly also assume $H_{0}=50$ km s$^{-1}$ Mpc$^{-1}$ and $q_{0}=0.5$). 
Our values of \n seem to lie approximately in the same range
as those obtained by Hill \& Lilly, although at the higher redshift end, Hill \& Lilly's
values for a few sources are larger.

%**************************************************************************

\section{Radio luminosity and redshift dependences of the clustering strength}
\label{section:section7}

In order to examine whether the strength of galaxy clustering around the quasars
is determined by epoch or radio luminosity, we looked for correlations between 
$B_{\rm gq}$, redshift and radio luminosity. 

\begin{figure}
\psfig{file=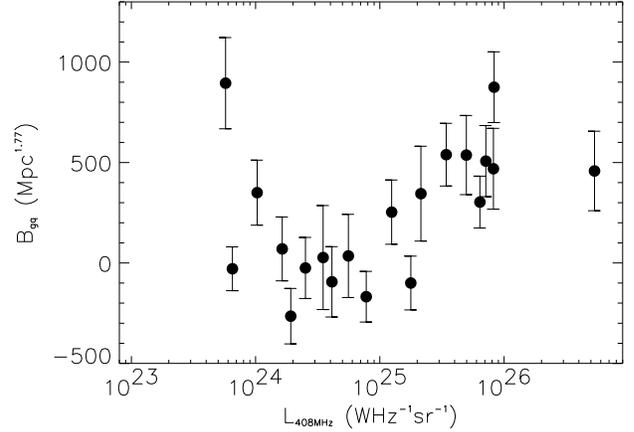}
\caption{\bgq plotted as a function of radio luminosity at 408 MHz.}
\label{fig:figure7}
\end{figure}

In Fig.~\ref{fig:figure7} we have plotted $B_{\rm gq}$ as a function of 
radio luminosity at 408 MHz, $L_{\rm 408MHz}$. The figure suggests that a 
positive correlation between \bgq and radio luminosity may be present,
at least there seems to be more of a trend with radio luminosity than with
redshift if we compare with Fig.~\ref{fig:figure4}.

Our sample is rather small and covers  
a relatively narrow redshift range, we therefore included the measurements of 
quasar environments down to $z\approx0.2$
from YG87, EYG91 and Yee \& Ellingson \shortcite{yee1}.
We used photometry from the NED to find flux densities and spectral indices to calculate 
radio luminosities for the literature sources.
In a few cases where the source was not available in NED, we used the NRAO VLA 
Sky Survey (NVSS) \cite{nvss}.
Since one of the aims was to investigate radio luminosity dependences,
we excluded sources that were flat-spectrum
quasars (nine sources in EYG91 and five from Yee \& Ellingson, as well as 
5C6.189 from our own sample), leaving us with a total of 71 radio-loud steep-spectrum quasars.
(The three sources in table~2 of EYG91 found to be radio-quiet were also omitted.)

\begin{figure}
\psfig{file=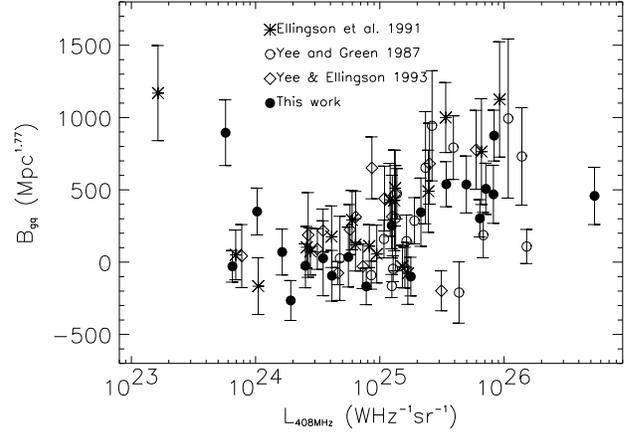}
\caption{\bgq versus radio luminosity for objects in our sample together with
steep-spectrum quasars from the literature.}
\label{fig:figure8}
\end{figure}

In Fig.~\ref{fig:figure8}, we have plotted the $B_{\rm gq}$'s for all sources 
as a function of radio luminosity. 
To examine whether there are any trends in the data,
we divided the quasars into redshift and luminosity sub-samples and computed 
the mean \bgq in each sub-sample. 
The result of this is shown in Table~\ref{table:table5}.

The mean \bgq for all quasar fields in our sample, 249$\pm75$, 
is seen to be consistent with the mean of 304$\pm$51 for the literature sources, and the mean of 
289$\pm$42 Mpc$^{1.77}$ for the combined samples.
The fact that the normalization of \bgq was done in a self-consistent manner,
both by us and the other investigators, is reassuring when we utilize their 
results in analyses like this. We also bear in mind the good agreement about
the richness of the MRC0405$-$123 field. As seen in Table~\ref{table:table5}, the
literature samples show a higher 
$\left<B_{\rm gq}\right>$ in the $z\geq0.5$ sub--sample than in the lower redshift
sub--sample, and we interpret this as the epoch-dependent evolution in \bgq claimed by
Yee \& Ellingson \shortcite{yee1}.
The mean \bgq for the combined samples is also seen to be larger in the high-redshift subsample, 
but here the difference smaller. 
However, the difference in the mean \bgq for 
the low and high radio luminosity sub--samples is more pronounced.
Here, we found
$\left<B_{\rm gq}\right>=136\pm72$ 
at low luminosities and $\left<B_{\rm gq}\right>=353\pm49$ Mpc$^{1.77}$ at high luminosities.
The sources in our own sample gave a similar result, but with an even larger difference,
$\left<B_{\rm gq}\right>=107\pm112$ at low luminosities and 365$\pm$89 Mpc$^{1.77}$ at 
high luminosities. The division between the high- and low radio luminosity sub--samples
was set at $L_{\rm 408MHz}=10^{24.75}$ W Hz$^{-1}$ sr$^{-1}$, which corresponds to the 
break in the radio luminosity function. 
  
In order to quantify the trends better, and to disentangle
the effects that redshift and radio luminosity may have on 
$B_{\rm gq}$, we followed the 
approach of Yates et al.\ \shortcite{yates} and computed the 
Spearman partial rank correlation coefficient, $r_{AX,Y}$, for the different samples.
The $r_{AX,Y}$ statistic is defined as 
\[ r_{AX,Y}\equiv \frac{r_{AX}-r_{XY}r_{YA}}{\left[\left(1-r_{XY}^{2}\right)\left(1-r_{YA}^{2}\right)\right]^{1/2}},\]
\noindent
where $r_{AX}$, $r_{XY}$ and $r_{YA}$ are the usual Spearman
rank correlation coefficients. The $r_{AX,Y}$ statistic is a number between
$-$1 and $+$1 and serves as an indicator of the trend of large values
of $A$ to be paired with large values of $X$ when the effect of the third
variable, $Y$, has been removed. A value of ($-$)$+$1 indicates a perfect 
(anti)correlation between $A$ and $X$ at constant $Y$.
For $N$ data points, the significance level of $r_{AX,Y}$ is
\[ D_{AX,Y}=\frac{1}{2}\left(N-4\right)^{1/2}\ln\left[\frac{1+r_{AX,Y}}{1-r_{AX,Y}}\right]\]
\noindent
\cite{macklin},
and is approximately normally distributed about zero with unit variance if 
the null hypothesis, that the $A$--$X$ correlation arises entirely from those of 
$Y$ with $A$ and $X$ separately, is true. 
Here, we let $A$, $X$ and $Y$ be combinations of $B_{\rm gq}$,
$L_{\rm 408MHz}$ and $z$, so that the correlation between e.g.\ 
\bgq and $L_{\rm 408MHz}$ with $z$ held constant can be found.
By using this statistic, we hope to determine whether there exists a correlation
in \bgq with either $z$ (independent of $L_{\rm 408MHz}$) or $L_{\rm 408MHz}$ (independent
of $z$).

\begin{table}
\caption{Mean \bgq for sub-samples of steep-spectrum quasars. The results for our sample,
the literature samples (YG87; EYG91; Yee \& Ellingson 1993) and all samples combined, are shown.
Radio luminosity, $L_{\rm 408MHz}$ is given in W Hz$^{-1}$ sr$^{-1}$ and the unit of \bgq is Mpc$^{1.77}$.
The number of objects in each sub--sample is denoted by $N$.}
\begin{tabular}{llllllll}
  & \multicolumn{2}{c}{} & \multicolumn{2}{c}{} & \multicolumn{2}{c}{Our +} \\
  & \multicolumn{2}{c}{Our} & \multicolumn{2}{c}{Literature} & \multicolumn{2}{c}{literature} \\
	      &                           &     &   & &                         &  \\        
Sub--sample              & $\left<B_{\rm gq}\right>$ & $N$ & $\left<B_{\rm gq}\right>$ & $N$ & $\left<B_{\rm gq}\right>$ & $N$ \\
             	      &                           &     &   & &                         &  \\        
All quasars                            & 249$\pm$75  & 20 & 304$\pm$51 & 51 & 289$\pm$42 & 71\\
$z<$0.5                                &             &  0 & 211$\pm$53 & 31  & 211$\pm$53 & 31\\
$z\geq$0.5                             & 249$\pm$75  & 20 & 450$\pm$94 & 20  & 349$\pm$61 & 40\\
$L_{\rm 408}<$10$^{24.75}$     & 107$\pm$112 &  9 & 158$\pm$97 & 12 & 136$\pm$72 & 21\\
$L_{\rm 408}\geq$10$^{24.75}$  & 365$\pm$89  & 11 & 349$\pm$59 & 39 & 353$\pm$49 & 50\\
\label{table:table5}
\end{tabular}
\end{table}

\begin{table}
\caption{Spearman partial rank correlation coefficients for the three different
samples. Our sample, samples from the literature (YG87; EYG91; Yee \& Ellingson 1993), and
our sample combined with the literature samples. The lower half of the table shows the
results of the correlation analysis when the compact steep-spectrum sources were 
excluded. The numbers in parenthesis behind the correlation coefficients are significances.}
\begin{tabular}{llrrr}
Sample & $N$ & $r_{B_{\rm gq}L_{\rm 408},z}$ & $r_{B_{\rm gq}z,L_{\rm 408}}$ & $r_{L_{\rm 408}z,B_{\rm gq}}$ \\
       &    &      &     &         \\
Our       & 20 & 0.41 (1.7$\sigma$) & 0.02 (0.1$\sigma$)& 0.22 (0.9$\sigma$) \\
Literature & 51 & 0.26 (1.8$\sigma$) & 0.19 (1.3$\sigma$) & 0.39 (2.8$\sigma$) \\
Combined  & 71 & 0.39 (3.4$\sigma$) & 0.10 (0.8$\sigma$) & 0.23 (1.9$\sigma$) \\
\hline
Our       & 14 & 0.60 (2.2$\sigma$) & $-$0.18 ($-$0.6$\sigma$) & 0.41 (1.4$\sigma$) \\
Literature &  44 & 0.27 (1.8$\sigma$) & 0.14 (0.9$\sigma$) & 0.39 (2.6$\sigma$) \\
Combined  & 58 & 0.47 (3.7$\sigma$) & 0.02 (0.1$\sigma$) & 0.22 (1.6$\sigma$) \\
\end{tabular}
\label{table:table6}
\end{table}

The Spearman partial rank correlation coefficients were computed in three different
cases: Our own sample, the literature samples (YG87+EYG91+Yee \& Ellingson)
and the samples combined.
The resulting correlation coefficients and their significances are shown in 
Table~\ref{table:table6}. 
The table is divided horizontally, where the upper part
shows the Spearman coefficients calculated by including all steep-spectrum quasars, whereas in the
lower part we have included only those steep-spectrum sources which
have extended radio morphologies, i.e.\ 
we have excluded quasars that have sizes $r<30$ kpc and are classified as
CSS sources.  The CSS sources may not sample the large-scale density distribution outside 
the ISM of the host galaxy, and therefore might not be expected to follow the same 
trends as the rest of the source population.
The Spearman statistic without the CSS sources remains almost unchanged, however.

The numbers in Table~\ref{table:table6} seem to indicate a weak, but significant,
dependence of \bgq on radio luminosity. Both in our sample and in the combined samples,
it is the most significant correlation, and it is
strongest when all the samples are combined, giving
$r_{B_{\rm gq}L_{408},z}=0.47$ with a 3.7$\sigma$ significance when the 
CSS sources are excluded. 
In the literature sample, it is the correlation between radio luminosity and
redshift that dominates. This makes it difficult to interpret any changes in 
$\left<B_{\rm gq}\right>$ within the two sub--samples, and especially we cannot be
certain that there is a true evolution with redshift.
However, both in our sample and in the combined samples $r_{B_{\rm gq}z,L_{408}}$ is negligible 
and insignificant compared to $r_{B_{\rm gq}L_{408},z}$.
The result of this analysis is therefore 
suggestive of a radio luminosity dependence in $B_{\rm gq}$, and we believe that 
there is no evidence for a redshift dependence 
as has been claimed by e.g.\ YG87 and Yee \& Ellingson \shortcite{yee1}.

We may consider that if our estimates of \bgq are systematically different
from the others, then a bias could be introduced, but we do not think that this is 
the case. First, the agreement about the value of \bgq for the field around 
MRC0405$-$123 is very good. Second, the trends within our own sample 
are the same as those found for the combined samples. Third, using 
the \bgq estimates determined from the raw counts in our data also give a 
significant correlation between \bgq and $L_{\rm 408MHz}$,
and produces the same trend as above when we combine with 
the literature sources. Fourth, the result of the correlation analysis using the 
recomputed $B_{\rm gq}$'s with EYG91's luminosity function as normalization
also show that the $B_{\rm gq}$--$L_{\rm 408MHz}$ correlation is the strongest. 
However, we may note, as mentioned by Yates et al.\ \shortcite{yates},
that $r_{B_{\rm gq}z,L_{408}}$ computed for a combination of samples 
that have been analysed by different authors should be viewed  
with some caution since differences in $\Phi\left(m_{\rm lim},z\right)$ 
and the assumed cosmology may introduce spurious effects.

%****************************************************************************************

\section{Discussion}
\label{section:section8}

\subsection{Comparison with radio galaxy environments}

Previously, studies of radio galaxies spanning a wide range in 
radio luminosity at a given redshift have found that the richness of radio 
galaxy environments is increasing with redshift, but 
does not correlate with radio luminosity (Hill \& Lilly 1991; Allington--Smith 
et al.\ 1993). 
It would be intriguing if a correlation between clustering strength and radio 
luminosity was found for radio-loud quasars but not for radio galaxies. 
The `Unified Scheme' for radio-loud AGN predicts that these are the same 
type of object viewed from different angles, so the richness of the 
environments should have the same behaviour with radio luminosity and redshift.
Why, then, are the radio quasars apparently different to the radio galaxies?

First, we consider the possibility that selection effects are responsible. 
Our sample was initially chosen such as to consist
of one low- and one high radio luminosity sub--sample.
The sources in these two sub--samples were selected randomly 
from the complete radio/optical flux-limited 7CQ \cite{riley} and MAQS samples
\cite{serjeant}, on the basis of redshift, right ascension and declination. 
The only possible source of 
bias would be a selection effect which leads us to give preference to
less dense environments at low radio luminosities (the 7CQ quasars) and more
dense environments at high radio luminosities (the MAQS quasars).

This selection effect is possible if the radio luminosity of a 
quasar is controlled {\em both}
by the environmental richness/density ($B_{\rm gq}$) and the bulk 
kinetic power in the radio jets (optical luminosity). 
We discuss this in greater detail in section~\ref{section:section83}.
The 7CQ sample has a relatively bright optical flux limit, which 
selects sources with unusually high optical luminosities for their radio luminosities.
The objects that we miss in the 7CQ survey are therefore optically faint with
low jet powers, so their radio luminosity
must be enhanced by a dense environment. We may therefore be biased against
optically faint quasars in dense environments and favour quasars in low-density 
environments instead, characterized by low $B_{\rm gq}$'s. 

In the MAQS sample, where the optical flux limit is faint, the effect would be 
to favour quasars in dense environments, i.e.\ high $B_{\rm gq}$'s. 
A faint quasar with weak jets 
in a dense environment can have the same radio luminosity
as an optically bright quasar in less dense 
environments since the powerful jets of the
latter will enhance the radio luminosity. Due to the steep increase of 
the optical quasar luminosity function towards low luminosities, we have 
more optically faint sources than we have bright sources, 
and since the sample also is radio flux-limited, 
we miss optically faint sources in low-density 
environments. The preference is thus given to optically faint sources 
in high-density environments having high $B_{\rm gq}$'s.

In the YG87 and EYG91
samples similar effects may operate, as most radio quasar samples are selected
with an optical flux limit close to the POSS--I plate limit, but they are
harder to quantify as the samples are not complete.

A second possible explanation is that we are not comparing like with like, 
especially at radio luminosities close to the break in the radio luminosity 
function. Recently, Laing et al.\ \shortcite{laing94} found evidence for two
different types of radio galaxies: low-excitation narrow-line radio galaxies 
(LEG), appearing as typical radio galaxies in radio,
but with weak optical emission lines that commonly have low
excitation, and high-excitation narrow-line radio galaxies (HEG) with strong, 
high-ionization emission lines. The HEG's are thought to be the objects 
which unify with quasars, whereas the LEG's, unless seen with their jets
pointing almost directly along the line of sight (when they are seen as 
BLLac's) are radio galaxies at all orientations since no broad lines are 
observed. If the 
clustering properties of HEG's and LEG's differ,
then a sample of radio galaxies with a mix of both HEG's and LEG's (such as 
that of Hill \& Lilly 1991) must 
be considered less uniform and difficult to compare with quasars. 

\subsection{The clustering properties of quasars and radio galaxies}

Because our quasars are typically in poor clusters, we can estimate 
their clustering (i.e.\ $B_{\rm qq}$) by assuming the ratio 
$B_{\rm qq}/B_{\rm qc} = B_{\rm cc}/B_{\rm cg} \approx 1.7$, where 
$B_{\rm cc}$ and $B_{\rm cg}$ are the cluster-cluster and cluster-galaxy 
correlation amplitudes respectively, from  Croft et
al.\ \shortcite{croft97} and Croft et al.\ (1999). This gives a
correlation amplitude for radio-loud quasars of $\approx 490$
Mpc$^{1.77}$ (after correcting all numbers to a $\gamma=1.77$ at 0.5 Mpc
for consistency). This is large compared to low redshift radio galaxies 
measured by Peacock \& Nicholson \shortcite{peacock} 
who obtain $B_{\rm g^*g^*} \approx 240$ Mpc$^{1.8}$, 
and the faint radio galaxies in the FIRST survey, for which 
Magliocchetti et al. \ \shortcite{magliocchetti} 
estimate $B_{\rm g^*g^*} \approx 220$ Mpc$^{1.8}$. 
Both the Peacock \& Nicholson and 
the Magliocchetti et al.\ studies are dominated by lower luminosity 
objects than in our quasar sample, further evidence of a radio luminosity 
and/or redshift dependence of clustering strength.
Also pertinent to this is the 2-degree Field (2dF) QSO Redshift Survey 
\cite{2df} currently underway at the 
Anglo-Australian Telescope, which will, in due course, measure the 
clustering of high-$z$ radio-quiet quasars directly, and also determine
the evolution in the clustering strength with redshift.

\subsection{The link between cluster richness and radio luminosity}
\label{section:section83}

Does our discovery of a correlation between radio luminosity and $B_{\rm gq}$
imply that environment is the primary factor in controlling radio luminosity?
If so there are at least three ways in which such a situation could come 
about:
(1) the environment determines the bulk kinetic power in the radio jets, 
$Q$, through a correlation of cluster richness 
with the black-hole mass in the central group/cluster galaxy; 
(2) the environment determines $Q$ through more AGN fuel being 
available in richer environments, or (3) $Q$ is independent of environment and 
similar for all radio sources, with the radio luminosity being determined
solely by the density into which the source expands.

The argument in favour of possibility (1) is based on 
the correlation between black hole mass and the mass of the spheroidal 
component suggested by Kormendy \& Richstone \shortcite{kr95} and Magorrian 
et al.\ \shortcite{magorrian}. This implies that the giant elliptical hosts 
of radio galaxies and quasars should have high black hole masses, $\sim 
10^{8}$--$10^{9} M_{\odot}$. If the jets are powered by accretion then 
it seems reasonable to expect that the maximum accretion rate is proportional 
to the Eddington rate, and hence the black hole mass. So more massive 
galaxies should power more luminous radio sources. These massive galaxies will always prefer richer 
environments, and the correlation between luminosity and cluster richness may 
just reflect an increasing mass of the host galaxy (and hence of the black 
hole). The principal argument against this is the lack of any obvious 
correlation of host galaxy luminosity with radio luminosity, at least out 
to $z\sim 0.6$ \cite{hstdata}, although there is a suggestion that 
by $z\sim 1$ the hosts of radio galaxies do show such a trend \cite{roche98}.

The possibility (2), that fuelling of the AGN determines its environment
is discussed by Ellingson, Green \& Yee \shortcite{ellingson2}. A 
group or poor cluster environment may be ideal for the fuelling of
a black hole, as encounters will be more common than in the field, but 
will be of low enough relative velocity to disrupt the interacting
galaxies and cause gas to flow into the centre. 
Lake et al.\ \shortcite{lake}, however, have also demonstated that galaxy 
harassment (frequent high-speed galaxy encounters) in relatively high 
velocity dispersion clusters may be an efficient mechanism for 
transporting gas to the centre of galaxies. If this mechanism is responsible
for fuelling the black hole it may explain the correlation of radio luminosity
with cluster richness, but cannot explain why low-power FRI radio sources are
also found in rich clusters.

Possibility (3) is suggested by 
models of radio sources which show that the radio luminosity 
of an expanding radio source is primarily
determined by the bulk kinetic power in the radio jet, $Q$, and the 
environmental density, $\rho$, as a function of distance from the 
nucleus, $r$, usually parameterised by
\[ \rho \propto \rho_{0} \left(a_0/r\right)^{\beta}, \]
\noindent
where $a_0$ and $\rho_0$ are the characteristic radius and density,
and $\beta$ is a power-law slope. 
Kaiser, Dennett--Thorpe \& Alexander (1997) show how the track of 
the expanding radio source on the radio-luminosity -- size 
diagram is determined by different values of these parameters.
The small dispersion in 
galaxy mass (and therefore perhaps black hole mass) of radio source
host galaxies might suggest that, if all radio sources are accreting matter
at the same fraction of the Eddington rate, the values of $Q$ should be 
similar for all radio sources. However,
it is clear from the models that radio luminosity is also strongly
dependent on the mean environmental density. This is because
objects located in dense gaseous haloes have lower losses due 
to adiabatic expansion, and the transfer of AGN power 
into radiation thus becomes very efficient due to the 
reduced expansion losses. Indeed, Barthel \& Arnaud \shortcite{barthelarnaud} 
use the case of Cygnus A, which is in such an environment, to argue
that environmental density {\em controls} radio source luminosity.
Consistent with this, Hutchings et al.\ \shortcite{hutchings} claim to 
find a weak, positive correlation between lobe luminosity of quasars 
and $B_{\rm gq}$. 

On the other hand, a relation between radio and [{\rm OIII}] emission-line
luminosity is observed (Rawlings \& Saunders 1991; Willott 
et al.\  1999). The [{\rm OIII}]\,$\lambda$5007 luminosity is believed 
to be powered by UV radiation from the nucleus (for a different view, 
see Dopita \& Sutherland 1995), and these observations thus argue that 
radio luminosity is primarily determined by jet kinetic power.
Serjeant et al.\ \shortcite{ser_raw} also find that the radio and optical
luminosity of quasars correlate. If both are
related to the luminosity of the AGN powered by accretion, then this also 
suggests that jet power, not environmental density, is the most important
factor in determining the radio luminosity. 

\begin{figure}
\psfig{file=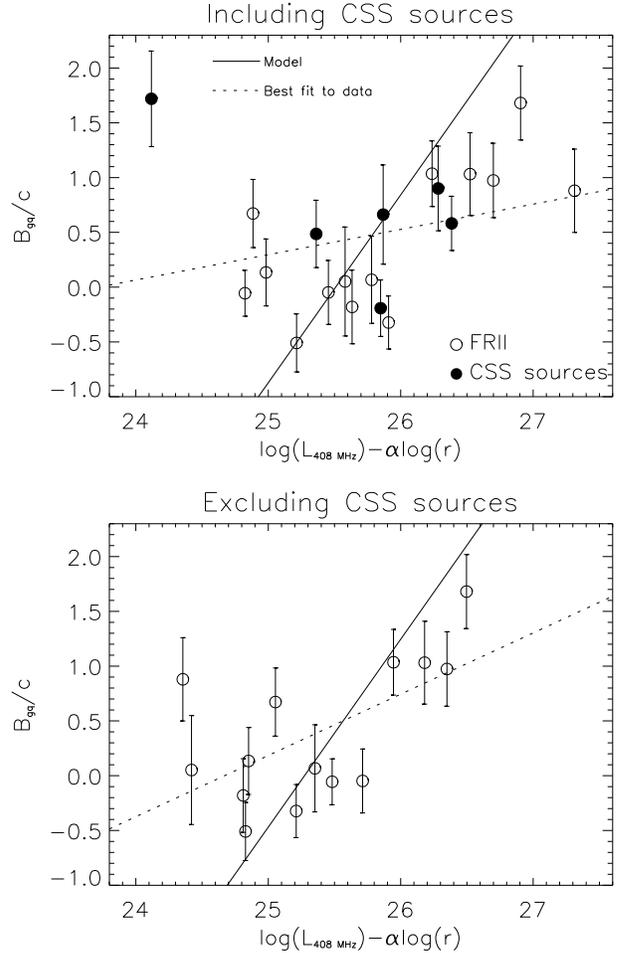}
\caption{The solid line corresponds to the predicted relationship between
quasar radio luminosity and environment from simple models assuming 
that environment is entirely responsible for determining the radio luminosity.
The predicted correlation is much stronger than what is observed, shown by
the dotted line. In this plot,
we have assumed that $\log\rho=B_{\rm gq}/c$, where $c=520.6$. The upper plot shows the
relationship when all quasars are included in the fit with a radio luminosity--size 
slope of $\alpha=-0.408$. In the lower plot, the CSS sources in the sample have
been ignored, and $\alpha=-0.239$. It may be noted that this rather simple model 
goes very quickly to negative \bgq values. This is implicit in the model due to the 
assumption that \bgq is proportional to $\log\rho$.}
\label{fig:figure9}
\end{figure}

To investigate the possible implications
of (3) for our study though, we have constructed a simple model of
how radio luminosity should relate to $B_{\rm gq}$.
We assumed that $B_{\rm gq}=c\log\rho$ and calibrated this 
relation using the gas 
densities from X-ray observations of 21 Abell clusters in Jones \& Forman 
\shortcite{jones} for which Yee \& L{\'o}pez--Cruz
\shortcite{yeelopez} have calculated $B_{\rm cg}$.
The median density and cluster richness for these 21 
clusters was found to be 3.03$\times$10$^{-3}$ cm$^{-3}$ and 1075 Mpc$^{1.77}$. 
We took this combination of $\rho$ and $B_{\rm cg}$ to be typical for 
rich clusters. For galaxies with no enhancement in the surrounding galaxy 
density, we used $B_{\rm gg}=60$ Mpc$^{1.77}$ \cite{davis} and a density 
of $3.4\times10^{-5}$ cm$^{-3}$, which is a typical gas density at 100 kpc 
in isolated ellipticals \cite{forman}. This gave a slope of $c=520.6$ for the
$B_{\rm gq}$--$\log\rho$ relation.

We then assumed that $L_{\rm 408MHz} \propto r^{\alpha}\rho^{7/12}$, as predicted 
by simple analytical models of radio sources (e.g.\ Miller, Rawlings \& 
Saunders 1993). The slope $\alpha$, was found by fitting a straight line
to the data in a $\log\L_{\rm 408MHz}$--$\log r$ plot, which came out as
$\alpha=-0.408$. For consistency, we also found the best-fitting slope
for the extended ($>30$ kpc) sources only, which gave $\alpha=-0.239$.
It is interesting to note that the quasars show an anticorrelation between 
radio luminosity and size, but that the population of radio sources as a whole 
does not. A likely explanation could be the optical selection effects in the 
7C sample. Also the emission at lower frequencies might be absorbed 
in the CSS sources, which would explain why there are fewer CSS sources among the
7C quasars (selected at 151 MHz) as opposed to the MRC quasars (selected at 408 MHz).
The correlation between radio luminosity and size also weakens when we exclude the
CSS sources.

By assuming a fixed jet power, $Q$, for all the sources, we then 
evaluated the relationship between $L_{\rm 408MHz}$ and \bgq as
\[ B_{\rm gq}/c=\frac{12}{7}\left(\log L_{\rm 408MHz}-\alpha\log r\right), \] 
\noindent 
in the two cases of $\alpha$.
The relation is plotted in Fig.~\ref{fig:figure9}, where the predicted slope
of 12/7 is seen to be steeper than what the data indicates.
So the correlation of \bgq with environment is weaker 
than we expect from simple models in which the density alone determines the radio luminosity.
We can therefore rule out as strong a 
$B_{\rm gq}$--$L_{\rm 408MHz}$ dependence as we would see if all objects had 
about the same jet power, and environment was entirely responsible for 
determining radio luminosity. 
Nevertheless, as the luminosity function for 
the radio jet power is likely to be steeply declining at high powers, it seems 
not unlikely that selection effects could operate to produce some correlation between 
$B_{\rm gq}$--$L_{\rm 408MHz}$ without it being as strong as it would be in this
rather extreme model in which the jet power is the same for all radio sources.

Given the large scatter in both the $B_{\rm gq}$--$L_{\rm 408MHz}$ and 
the $L_{\rm 408MHz}$--$M_{B}$ (or $L_{\rm [{\rm OIII}]}$) correlations
it is quite possible that both environment and radio jet power play important r\^{o}les
in determining the radio luminosity. The relationship 
between radio sources and their environments must be complex, and the vast majority
of radio sources may lie in some sort of cluster-like environment, from groups of 
only a few galaxies to clusters as rich as Abell class 1 or more.
The weak correlation we see can probably be 
explained if we assume that radio sources occupy a range of environments from fairly
isolated ellipticals to rich clusters, and their radio luminosities are 
determined both by the environment and the kinetic power in the jets.

%***********************************************************************

\section{Summary}
\label{section:section9}

We have investigated the galaxy environments around 21 radio-loud,
steep-spectrum quasars in the luminosity range 
$23.8 \leq \log\left(L_{\rm 408MHz}/{\rm W Hz^{-1} sr^{-1}}\right) \leq 26.7$ 
and redshift range
$0.5 \leq z \leq 0.82$, and find an excess of galaxies within a 0.5 Mpc radius
centered on the quasar. The galaxy excess in each quasar field was quantified
using the galaxy--quasar cross--correlation amplitude, $B_{\rm gq}$,
and an Abell type measurement, $N_{0.5}$. 
Our estimates of \bgq is seen to fall approximately in the same range
as that found for quasar environments by other investigators. 
We also find that $N_{0.5}$, although
originally intended for fields around radio galaxies \cite{hill}, is 
applicable to quasars fields too, and that $N_{0.5}$ for our quasar fields
is consistent with \n for fields
around radio galaxies studied by Hill \& Lilly \shortcite{hill}.

The average \bgq of our quasar sample was found to be 265$\pm$74 Mpc$^{1.77}$, 
or in terms of $r_{0}$, the correlation length of the galaxy--quasar cross--correlation 
function, $r_{0}=23.4$ Mpc.
If the galaxies around the quasars were distributed as low-redshift
field galaxies, \bgq would be $\approx60$ Mpc$^{1.77}$ and 
$r_0 \approx 10$ Mpc as found for the galaxy-galaxy correlation function 
in the field (e.g. Davis \& Peebles 1983). This shows that the quasars are 
sited in richer than average galaxy environments, and two of the quasar 
fields have \bgq above 740 Mpc$^{1.77}$, considered here to be characteristic for clusters with
Abell richness class 1 or greater. 

Our sample extends previous quasar samples to higher redshifts whilst maintaining
the same range in radio luminosity. We compared with \bgq measurements in fields
around radio-loud quasars from these samples (YG87; EYG91; Yee \& Ellingson 1993), and 
detected a weak, but significant, correlation between radio luminosity and environment
richness. The correlation between \bgq and environment was found to be much weaker
than expected from simple models in which gas density alone determines 
the radio luminosity. 
Probably, the correlations between \bgq and environment, and 
between jet power, $Q$, and quasar UV/optical luminosity can be explained if both 
$Q$ and environment play significant roles in determining the luminosity 
of a radio source.

Studies of radio galaxies (e.g.\ Hill \& Lilly 1991) have found that
environmental richness increases with redshift and that there is no correlation 
with radio luminosity, contrary to our observations of radio-loud quasars.
The apparent difference between radio-loud quasars and galaxies would cause 
problems for the unified scheme of radio-loud AGN unless the difference 
can be explained by selection effects or a mix of two
different types of objects \cite{laing94} in radio galaxy samples. 

Our main conclusion is that the radio-loud steep-spectrum quasars in our sample on 
average occupy environments typical of poor clusters with Abell richness 
class 0, but that quasar environments may cover a wide range in richness, 
from groups of galaxies and poor clusters to clusters as rich as Abell class 1 or more.
In a conventional dark matter cosmology, very rich clusters at redshifts
0.5--0.8 should be rare as they would not have had time to form.
Nevertheless, a few very luminous $z\sim 0.8-1.3$ X-ray selected clusters have 
been found (e.g.\ Henry et al.\ 1997; Rosati et al.\ 1999), 
so at least some 
very gas-rich environments exist at high redshift. It is thus perhaps 
surprising that so few radio-loud objects are found in clusters of Abell 
richness 1 or higher, given the arguments about both galaxy--black hole mass 
and the relationship between gas density and radio luminosity. This adds weight
to the suggestion that fuelling problems may be the reason why radio sources
seem to prefer only moderately rich environments.

%*******************************************************************

\section*{Acknowledgments}

We are grateful to the staff at the NOT and the McDonald Observatory
for their help with the observations and to D.~Burstein for kindly
providing maps of galactic extinction in electronic form. We also thank
S.~Rawlings for communicating redshifts for the 7CQ survey prior to 
publication and K.~Blundell for the sizes of two of the MRC
quasars.

ML and MW were guest observers at the McDonald Observatory. The
Nordic Optical Telescope is operated on the island of La~Palma
jointly by Denmark, Finland, Iceland, Norway, and Sweden, in the
Spanish Observatorio del Roque de los Muchachos of the Instituto de 
Astrofisica de Canarias.
We thank the British Research Council and the Research Council of Norway for a 
joint travel grant, and MW 
acknowledges further travel support from the Research Council of Norway.

This research is partially based on observations made with the NASA/ESA Hubble Space Telescope,
obtained at the Space Telescope Science Institute.
STScI is operated by the Association of Universities for Research in 
Astronomy, Inc. under NASA contract NAS 5-26555.
{\sc iraf} is distributed by the National Optical 
Astronomy Observatories, which are operated by the Association of 
Universities for Research in Astronomy, Inc., under cooperative agreement 
with the National Science Foundation. 
This research has also made use of the NASA/IPAC extragalactic database
(NED) which is operated by the Jet Propulsion Laboratory, California
Institute of Technology, under contract with the National Aeronautics and 
Space Administration.

\end{document}